\documentclass[superscriptaddress,prx,longbibliography,twocolumn, footinbib]{revtex4-2}

\usepackage[dvipsnames]{xcolor}
\usepackage[space]{grffile}
\usepackage{amsmath, amssymb, amsfonts,bm}

\usepackage{latexsym}   
\usepackage{bbm} 
\usepackage{mathtools}  
\usepackage{placeins}   
\usepackage{comment}
\usepackage{float}

\newcommand{\addKN}[1]{\textcolor{blue}{#1}}

\begin{document}


\title{Magnonic Superconductivity}




\author{Khachatur G. Nazaryan$^{1}$, Liang Fu$^{1,*}$\\
  \small $^1$Department of Physics, Massachusetts Institute of Technology, Cambridge, MA 02139\\
  \small *Corresponding author: liangfu@mit.edu (L.F.)
}

\date{\today}

\begin{abstract}

We uncover a superconducting state with partial spin polarization induced by a magnetic field. This state, which we call "magnonic superconductor", lacks a conventional pairing order parameter, but is characterized instead by a composite order parameter 
that represents the binding of electron pairs and magnons. We rigorously demonstrate the existence of magnonic superconductivity with high transition temperature in one- and two-dimensional Hubbard models with repulsive interaction. We further show that magnonic Cooper pairs can attract to form higher-charge bound states, which can give rise to charge-$6e$ superconductivity.

\end{abstract}

\maketitle


\section*{Introduction} 

Unraveling the mechanism of high-temperature superconductivity in doped cuprates and other correlated insulators is a long-standing challenge in condensed matter physics \cite{Lee2006, Norman2011}. 
An appealing idea is the resonating valence bond (RVB) theory  \cite{Anderson1973, Fazekas1974}, originally proposed for parent Mott insulators that are quantum spin liquids formed from highly entangled spin-singlet pairs. 
More recently, unconventional superconductivity has been discovered in a variety of two-dimensional (2D) materials, which emerges from doping insulating states. These include ZrNCl \cite{Nakagawa2021}, WTe$_2$ \cite{Fatemi2018}, 
and crystalline graphene multilayers \cite{Zhou2021,Zhou2022}. 
While undoped cuprates are antiferromagnetic Mott insulators, 
the parent states in these 2D materials are of a different nature. Undoped ZrNCl,  WTe$_2$ and crystalline graphite are all band insulators with an even number of electrons per unit cell. Upon doping, the insulating state in ZrNCl and WTe$_2$ evolves directly into the superconducting state. 
Unconventional superconductivity is evidenced by doping-induced BEC/BCS crossover in ZrNCl and strong violation of Pauli limit in WTe$_2$.

The emergence of unconventional superconductivity from doping band insulators in 2D materials opens up fresh perspectives on the pairing mechanism \cite{Dong2023,Cea2022,Chou2022, Ghazaryan2021,Chatterjee2022, Sous2023}. An interesting possibility is that two doped charges form a bound state in the insulating background. Counter-intuitively, recent works have shown that such two-particle bound state---often with non-$s$-wave pairing symmetry---can be formed due to interband fluctuations in band insulators with {\it repulsive} electron-electron interaction \cite{Slagle2020, Crepel2021, Crepel2022, Crepel2022_2,Crepel2023, vonMilczewski2023, Homeier2024, He2023}. This mechanism of electron attraction from repulsion gives rise to unconventional superconductivity upon doping.





In this work, we demonstrate an unconventional 
pairing and superconductivity that emerges from doping spin-polarized insulators  under a magnetic field.  
In our theory superconductivity is formed from the condensation of charge-$2e$ multiparticle bound states made of two doped charges and one or more magnons (spin-flips), which carry spin $S \geq 2$ distinct from Cooper pairs. 
We call such bound state a "magnonic Cooper pair'', and the corresponding superconductor a ``magnonic superconductor''.  Importantly, a magnonic superconductor does not have a standard pairing order parameter $\langle c^\dagger_{i\alpha} c^\dagger_{j \beta} \rangle = 0$, but is characterized by a composite order parameter, for example, $\langle c^\dagger_{i\uparrow} c^\dagger_{j\uparrow} c^\dagger_{k \uparrow} c_{l \downarrow} \rangle \neq 0$.

We demonstrate the formation of two-holes-one-magnon (2H1M) bound states in a class of strongly repulsive Hubbard models under a range of Zeeman fields, and identify the common mechanism. When the tight-binding energy dispersion exhibits three degenerate maxima, two doped holes are attracted to one magnon to form a three-body bound state with charge $2e$, i.e., a magnonic Cooper pair. This bound state is characterized by a large binding energy comparable to the hopping amplitude and a small size of a few lattice constants. The condensation of these spin-2 charge-2e bound states results in magnonic superconductivity.

To illustrate this kinetic mechanism for magnonic Cooper pair, we present concrete examples in both one- and two-dimensional Hubbard models with infinite-$U$. Starting with the fully spin-polarized insulator at half filling, we study
the system doped with two holes as a function of the magnetic field and find numerically exact solution for the
ground state. Below a critical field, the system transitions into a non-fully polarized state with one or more
magnons present. In both models, a robust 2H1M bound state is formed when the tight-binding energy dispersion satisfies the three-valley condition.

We note that related ideas of composite pair superconductivity have been considered in Kondo lattice systems \cite{Abrahams1995}, where spin-$0$ Cooper pair may couple to a spin-$1$ magnon to form a spin-$1$ charge-$2e$ order parameter. This is different from our case of spin-$2$ composite pairing.  Most closely related to our work is Ref.\cite{Batista2018}, which found similar composite pairing in a different 
model, albeit only in a very narrow range of magnetic field. Our work goes beyond specific model studies to identify a common mechanism for 2H1M bound state based on the kinetic energy dispersion. Finally, by analyzing the interactions between these 2H1M charge-$2e$ bosons, we 
demonstrate the formation of three-boson bound states in a certain parameter regime, which gives rise to charge-$6e$ superconductivity.

\textit{Triangular lattice Hubbard model under magnetic field.} For concreteness, we study the Hubbard model on the triangular lattice with nearest and next nearest neighbor hoppings $t_1$ and $t_2$, subject to a magnetic field $h$ that couples to electron spin: 
\begin{align}
    H= &-t_1 \sum_{\sigma; \langle \textbf{i}, \textbf{j} \rangle }\left(c_{ \textbf{i} \sigma}^{\dagger}c_{ \textbf{j} \sigma}+ h.c.\right)-t_2 \sum_{\sigma; \langle\langle \textbf{i}, \textbf{j} \rangle\rangle }\left(c_{ \textbf{i} \sigma}^{\dagger}c_{ \textbf{j} \sigma}+ h.c.\right)\nonumber\\
    &+U\sum_{ \textbf{i} }n_{ \textbf{i} \uparrow}n_{ \textbf{i} \downarrow} -\frac{h}{2}\sum_{ \textbf{i} }\left(n_{ \textbf{i} \uparrow}-n_{ \textbf{i} \downarrow}\right).
\end{align}
As we will show later, the essential physics of magnonic Cooper pairing  remains robust against additional perturbations such as longer-range hopping or interaction. 

\begin{figure}[t]
\begin{centering}
\includegraphics[width=0.9\columnwidth]{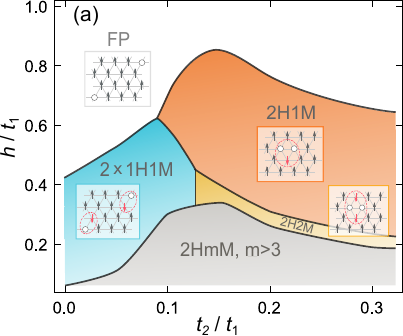}
\par\end{centering}
\caption{\textbf{Phase diagram of the infinite-$U$ Hubbard model}. Studied is the triangular lattice at half-filling with two doped holes, as a function of the magnetic field $h$ and tight-binding parameters $t_2/t_1$. Various multi-particle bound states of holes and magnons are found below the saturation field.  
}
\label{fig:Phase_Diagram}
\end{figure}

Exactly at half-filling (with one electron per site), the ground state  is a fully spin-polarized insulator above a critical field denoted as $h_J$. In the strong-coupling regime (large $U$), $h_J$ is determined by the spin exchange interaction $J_{a} = 4 t^2_{a}/U$ ($a=1,2$) and therefore is small. When one or more holes are introduced, however, a much larger saturation field (denoted as $h_1$) comparable to the hopping energy scale is required to fully polarize {\it all} the spins in the system. This is because the kinetic motion of holes is affected by the presence of magnons. This kinetic mechanism 
also leads to various bound states of holes and magnons, as we show below.        

We first review the case of one doped hole. Starting with the seminal work of Zhang, Zhu and Batista \cite{Batista2018}, the problem of one doped hole in two-dimensional triangular lattice Hubbard model has been thoroughly studied in recent years \cite{Davydova2023, Morera2023}. Below a critical field, the ground state with one hole is a non-fully polarized state that contains a spin polaron, a one-hole-one-magnon bound state with spin $S=3/2$ and a large binding energy on the order of the hopping amplitude.  

For the triangular lattice Hubbard model with only nearest-neighbor hopping ($t_2=0$), the spin polaron is shown to be the lowest-energy-per-charge excitation over a wide range of field. Consequently, the ground state at finite doping density is a spin polaron liquid that exhibits a magnetization plateau \cite{Davydova2023, Zhang2023}, which has been experimentally observed in semiconductor moir\'e superlattices \cite{Tao2024}.

For the purpose of studying pairing and superconductivity, this work is concerned with the system of two doped holes. Importantly, we consider the Hubbard model with both nearest and second nearest neighbor hoppings. As we will show below, $t_2\neq 0$ plays a crucial role in allowing for the binding of two holes and magnons.   

\begin{figure}[t]
\begin{centering}
\includegraphics[width=0.9\columnwidth]{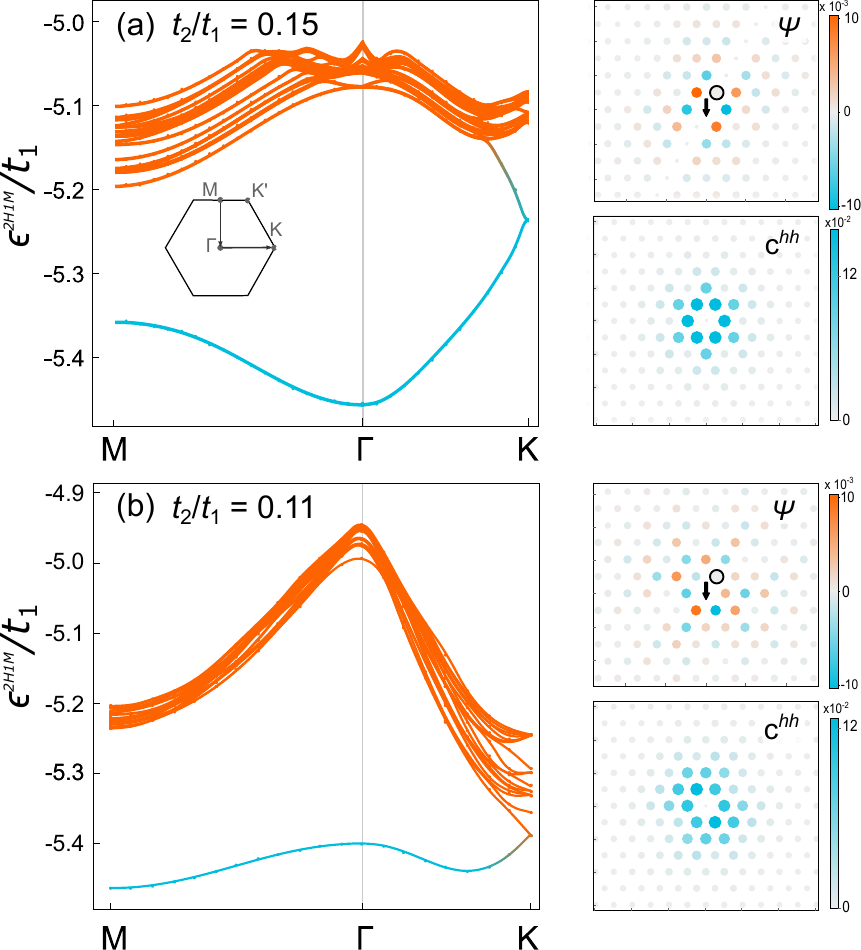}
\par\end{centering}
\caption{\textbf{Properties of the magnonic Cooper pair particles.} (a), (b) Many-body spectrum of the infinite-$U$ Hubbard model in the presence of two doped holes and one spin flip as a function of COM momentum along $M$-$\Gamma$-$K$ direction for $t_2/t_1 = 0.15$ and $0.11$, respectively. The lowest energy branch (blue color) corresponds to the two-holes-one-magnon bound state, which is separated from the high-energy continuum (orange color) by a large gap.  Also shown on the right are the ground state wavefunction at fixed positions of the spin-flip (down arrow) and the hole (white circle), and the hole-hole correlation function $c^{hh}$. 
}
\label{fig:Band_Structure}
\end{figure}

\section*{Results}

We use the exact diagonalization method to calculate the energy spectrum of the system with two holes and $m$ magnons, denoted as $2HmM$. 
First, we diagonalize the many-body Hamiltonian $H$ at zero field $h=0$ to find the lowest energy state with a fixed number of magnons, whose energy is denoted as $\epsilon^{2HmM}$. By utilizing the center-of-mass momentum as a good quantum number, we obtain results for systems up to $N = 36^2 = 1296$ sites ($m=1$), $N=6 \times 12=72$ sites ($m=2$), and $N=6 \times 6=36$ sites ($m=3$). In the presence of magnetic field, the energies of states with two holes and $m$ magnons are given by 
\begin{align}
    &E^{2HmM} = \epsilon^{2HmM} +(m+1) h , \label{eq:E2HmM} 
\end{align}
where the second term comes from the Zeeman energy cost associated with holes and spin-flips.

\begin{figure}[t]
\begin{centering}
\includegraphics[width=\columnwidth]{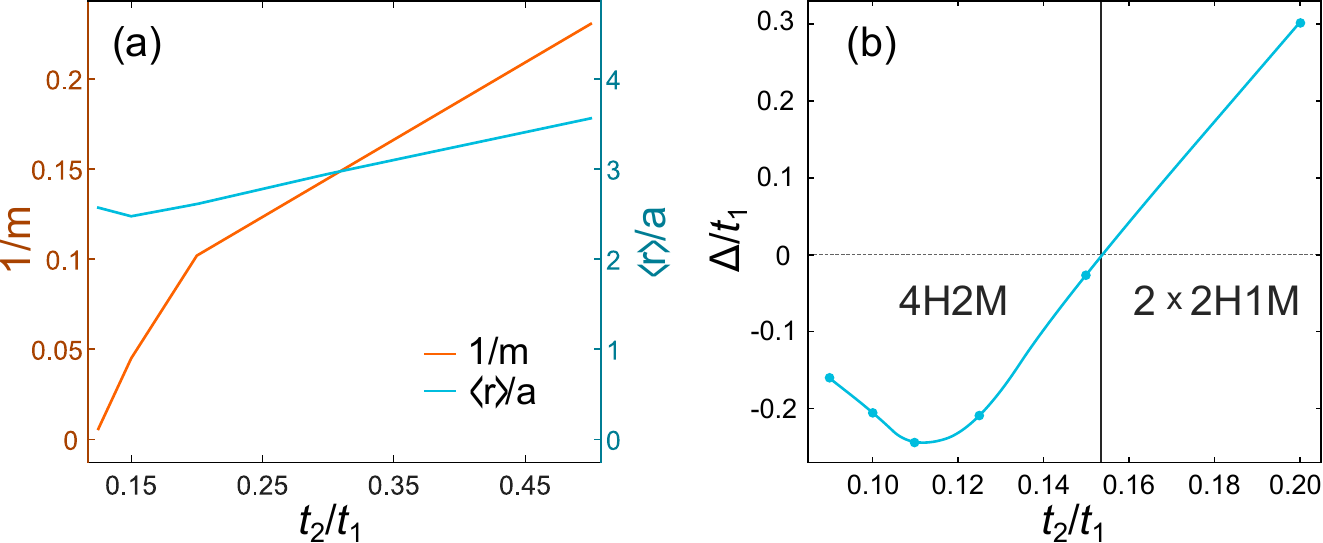}
\par\end{centering}
\caption{\textbf{Effective mass and interaction between magnonic Cooper pairs.} (a) The inverse effective mass in units of $t_1 a^2/\hbar^2$ (red) and average distance between holes (blue) of the magnonic Cooper pair (2H1M). The average distance was calculated on lattices $12\times 12$, $18\times 12$ and $18\times 18$, showing no sensitive dependence on the size, indicating a well-localized state. (b) Energy gap of 4H2M state vs 2 $\times$ 2H1M. For $t_2/t_1$ within $(0.09, 0.15)$ the magnonic Cooper pairs attract and form a composite particles, while for larger $t_2/t_1$ they repel.} 
\label{fig:Eff_mass_Av_Dist}
\end{figure}

For a given magnetic field $h$, by comparing these energies of different states, we determine the number of magnons $m$ which minimizes the energy. The full phase diagram of the ground state with two holes thus obtained is presented in the Fig. \ref{fig:Phase_Diagram}, as a function of $t_2/t_1$ and the magnetic field $h$.  
As the magnetic field is reduced, distinct hole-magnon bound states with different numbers of magnons are found. For a wide range of field below the saturation field denoted as $h_2$, the ground state of two holes is either two unbound spin polarons (1H1M) or a 2H1M bound state, depending on the tight binding model parameters. Below a lower critical field, bound states with two or more magnons appear. Below we will analyze various regions in detail.

Our findings reveal that for  small enough $t_2/t_1$, below the saturation field the fully polarized state (FP) 
transitions into a state with two separate charge-$e$ spin polarons, consistent with the previous works \cite{Davydova2023, Zhang2023}. In contrast, a small second-neighbor hopping $t_2\geq0.09 t_1$ is enough to drive the system into a different phase: over a wide range of field, we find a bound state of two holes and one magnon ($2H1M$). This establishes the formation of a  \textit{magnonic Cooper pair}, a composite boson of charge $2e$ and spin $S=2$.


We further calculate the many-body energy spectra of the system with two holes and one magnon as a function of the many-body momentum, shown in Fig.2. For two representative values of $t_2/t_1=0.11$ and $0.15$, the lowest energy branch is well separated from the the continuum by a gap.  
This branch corresponds to the energy dispersion of the $2H1M$ bound state as a function of its center-of-mass (COM) momentum. Note that when $t_2/t_1$ surpasses $0.15$, the ground state has COM momentum at $\Gamma$, while within the region $0.09< t_2/t_1 < 1/8$, the COM momentum shifts to $M$ point. The binding energy is remarkably large and on the order of the hopping energy. This implies that the binding of two holes is robust against perturbations to the model Hamiltonian, e.g., longer-range hoppings or electron-electron repulsions.


The internal structure of the $2H1M$ bound state can be visualized by examining its wavefunction, which depends on the positions of two holes and one magnon. By fixing the magnon at the origin and one of the holes at a neighboring site, the dependence of the wavefunction on the coordinate of the second hole is depicted in Fig.2. 
Additionally, the hole-hole correlation function  
$c^{hh} (\textbf{r}) = \langle n_h(\textbf{0}) n_h(\textbf{r} ) \rangle$ 
clearly shows that the two holes are tightly bound.  
Remarkably, the bound state size as defined by the average distance between the two constituent holes is only around $3$ lattice constants over a wide range of tight binding parameters, see Fig. \ref{fig:Eff_mass_Av_Dist}.




\begin{figure}[t]
\begin{centering}
\includegraphics[width=0.9\columnwidth]{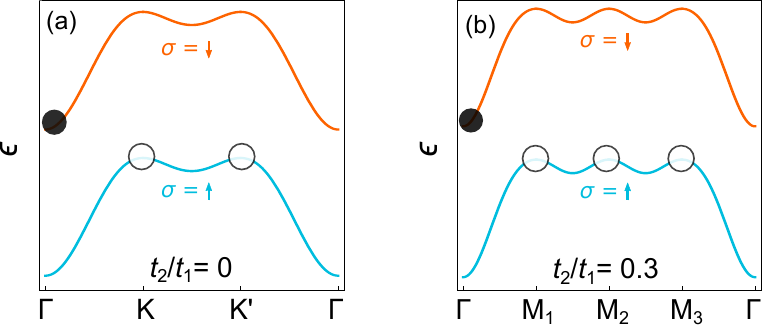}
\par\end{centering}
\caption{ 
\textbf{Three-fold degenerate band maxima as the key mechanism for magnonic Cooper pair formation.} (a), (b) Spin up and spin down electron dispersions in non-interacting limit (separated by Zeeman energy) for $t_2/t_1=0, 0.3$, respectively. The electron band minimum is at $\Gamma$ point. Meanwhile, the band maximum is double-degenerate at $K, K^\prime$ points for $t_2/t_1 < 1/8$; it becomes triple-degenerate for $t_2/t_1 > 1/8$ shifting to $M$ points. In (a) two holes are efficiently placed at band maxima allowing for polaron formation. In (b), system can host three holes at band maxima, leading to the formation of a magnonic Cooper pair.
}
\label{fig:Electron_disp}
\end{figure}





Insights into the origin of 2H1M bound state  can be gained from considering the kinetic energy dispersion in the non-interacting regime, illustrated in the Fig. \ref{fig:Electron_disp}. We observe that while the band minimum occurs at the $\Gamma$, the location of the band maximum highly depends on the parameter $t_2/t_1$. For weak $t_2$, the maximum is doubly degenerate, occurring at $K$ and $K^\prime$. However, when $t_2/t_1$ exceeds 1/8, the maximum shifts to $M$ points leading to a three-fold valley degeneracy. 
As a result, the system can host two holes in any of the 3 band maxima. In the absence of any magnons (i.e. the fully polarized phase), the energy of two-hole state in the Hubbard model is simply $2\epsilon_M$, with $\epsilon_M$ being the energy of a hole at $M$ point.
In contrast, in the presence of a magnon the translation symmetry of the two holes in the COM frame is broken, and the state of two holes can be in a coherent superposition of three valleys of the form: $|M_1,M_2\rangle + |M_2,M_3\rangle +|M_3,M_1\rangle$. 
And such coherent superposition lowers the energy of the system.
This picture also applies to the small $U$ limit. Here 2H1M state is a bound state of 3 spin $\downarrow$ holes and 1 spin $\uparrow$ electron. We have numerically verified that the transition to such 4 particle bound state occurs precisely at $t_2/t_1=1/8$, and the results are presented in the supplementary materials.

As the magnetic field is further reduced, the system continues to reveal its richness. Notably, when $t_2/t_1$ surpasses $0.13$, the phase characterized by two separate polarons (2 x 1H1M) transitions into a state where the polarons combine into a composite bound state (2H2M). In this phase, the charge carriers are bosons with a charge of 2$e$ and spin $S=3$. By further decreasing magnetic field, the system transitions into states with an increasing number of magnons. Figure \ref{fig:Phase_Diagram} shows that the 2H2M phase has a substantially narrower width compared to the 2H1M phase. As the field further decreases, we anticipate additional transitions to states with a larger number of magnons, eventually leading to magnon condensation.

\section*{Discussion}

Having established the phase diagram of two holes, we now discuss the effects of finite doping density. 
The existence of magnonic Cooper pairs over a wide range of magnetic fields naturally suggests the possibility of superconductivity. At low density our system maps to an interacting Bose gas where the boson corresponds to 2HmM bound states. If the interaction between these bosons is repulsive, the ground state of the 2D Bose gas will be a superfluid. 

We now determine the sign of interaction of the magnonic Cooper pairs. We perform ED study for a system with 4 holes and 2 magnons on a $6\times 6$ lattice and compare the energy with that of two separate 2H1M bound states. The Fig. \ref{fig:Eff_mass_Av_Dist}b) reveals that the interaction is attractive for $t_2/t_1$ within $(0.09, 0.15)$ and then shifts to repulsive for larger $t_2/t_1$.

Therefore, for $t_2/t_1 > 0.15$, the ground state at low density is a superconductor formed by Bose condensation of 2H1M magnonic Cooper pairs.  
It is well known that  the BKT superfluid transition temperature for a two dimensional Bose gas with weak repulsive interactions can be estimated as \cite{Prokofev2001, Roy2021}, $k_B  T_{\mathrm{BKT}}=\frac{n^\text{b}}{m}\frac{2 \pi \hbar^2}{\ln \left(380 \hbar^2 / m U_0\right)}$. 
Here $n^\text{b}=n/2$ is the boson density and $n$ the hole doping density, $m$ the effective mass of bosons, $k_B$ is the Boltzmann constant and $U_0$ is the contact repulsion strength between the bosons in the continuum. For our Hubbard model $U_0$ can be extracted from the microscopic parameters, and we expect it to be of order of hopping amplitude up to a numerical prefactor. But because the dependence on $U_0$ is through a $\ln$, the exact value should not affect the order of magnitude estimate of $T_{BKT}$. The important dependence is on the effective mass $m$ of the  2H1M state, which we determine from the many-body spectrum. The results on the Fig. \ref{fig:Eff_mass_Av_Dist}a) show that for $t_2>0.15 t_1$ the inverse mass $1/m$ is of order $0.1 t_1 a^2/\hbar^2$. Therefore, the superconducting transition temperature $k_B T_{BKT}$ is of the order of $0.1 t_1 \cdot\nu /\sqrt{3}$ where $\nu=n a^2\sqrt{3}/2$ is the number of holes per site. 
We compare $T_{BKT}$ with the Fermi energy, $E_F$, of the hole system in the fully polarized state, which for low doping can be found as $E_F = 2\pi/3 \cdot t_1\nu \sqrt{(9t_2/t_1-1)(1-t_2/t_1)}$. Here the factor $1/3$ comes from the three-fold degeneracy of the band minima ($M$ points). At a ratio of $t_2/t_1=0.2$, $E_F$ is approximately $1.6 \nu t_1$. Consequently, the superconducting transition temperature $T_{BKT}$ is approximately $5\%$ of $E_F/k_B$.

For $t_2/t_1 <0.15$, our ED calculation indicates that the two 2H1M bosons attract, yielding three degenerate ground states with COM momentum at $M$ points, separated by a finite energy gap to higher energy states.
We attribute this to the fact that the ground state is a bound state of two 2H1M bosons in different valleys, and therefore carries a total momentum $M_1+M_2=M_3$ etc. 
The presence of attraction between charge-$2e$ 2H1M bosons leads us to conjecture that three magnonic Cooper pairs with total momentum at $M_1 + M_2 +M_3 = \Gamma$ will form a stable composite bound state. This will give rise to a magnonic superconductivity through a Bose condensation of charge $6e$ and spin $S=6$ bosons. 

To explicitly illustrate this interesting scenario, we study an effective model of dilute boson gas with three flavors corresponding to the three $M$ valleys. The Hamiltonian for this model takes the following form dictated by symmetry:  
\begin{align}
&\mathcal{H}=\frac{1}{2}\sum_{\alpha,ij}\int d\textbf{r} \; m_{\alpha,ij}^{-1} \psi_{\alpha}^{\dagger} \boldsymbol{\partial}_{i}\boldsymbol{\partial}_{j}\psi_{\alpha} \nonumber\\
&+\frac{1}{2}\int d\boldsymbol{r} \; \bigg(\sum_{\alpha} U (\psi_{\alpha}^{\dagger})^{2}(\psi_{\alpha})^{2} -\sum_{\alpha\neq\beta} V n_{\alpha}n_{\beta}\bigg)\nonumber\\
&+\frac{1}{2}\int d\boldsymbol{r} \sum_{\alpha\neq\beta} J (\psi_{\beta}^{\dagger})^{2}(\psi_{\alpha})^{2}
    \label{eq:Effective_FT}
\end{align}
Here $\alpha, \beta = 1,2,3$ denote the different $M$ valleys; the first term describes the kinetic energy with anisotropic mass $m_{\alpha,ij}$ around the $M_\alpha$ point. The second and third terms represent the intravalley and intervalley density-density interactions. 
The third term is allowed by symmetry and results in the scattering of two particles from one valley  to another valley. In the low density limit, only contact interactions are relevant and their strengths are denoted as $U,V$ and $J$ (which are determined by our microscopic Hubbard model parameters). Importantly, the intervalley interaction $V$ is attractive and responsible for the binding energy of two 2H1M bosons from different valleys.       
We further analyze the effective model of bosons; details can be found in the Supplementary Material. To determine what kind of boson bound states is energetically favorable, we compare the binding energy per particle for  two-boson and three-boson states. Assuming that intravalley repulsion $U$ is repulsive and large, we find that the three-boson bound states are more favorable than the two-boson ones and even infinitesimal attraction $V<0$ is enough to form such bound states. 

Therefore, we conclude that over a wide range of magnetic field, our Hubbard model at low density is a dilute system of magnonic Cooper pairs, charge-$2e$ bosons consisting of two holes and one magnon. The boson energy dispersion may have a unique minimum at $\Gamma$ or three degenerate minima at $M$. In the former case the system is a charge-$2e$ magnonic  superconductor, while in the latter case the system exhibits charge-$6e$ superconductivity due to boson triple condensation. Both magnonic superconductors are characterized by composite order parameters involving both electrons and magnons. Last but not the least,  the binding between hole pairs and magnons into magnonic Cooper pairs implies that the magnetization as a function of the magnetic field exhibits a plateau where the total spin $S$ depends only on the doping $\delta$: 
\begin{eqnarray}
S = N(1 - 2\delta)/2    
\end{eqnarray} 
with $N$ is the number of unit cells. This value differs from the fully spin-polarized state which has a total spin $N(1 - \delta)/2$ as well as the spin-polaron metal, where the total spin is $N(1 - 3\delta)/2$ \cite{Zhang2023}.

\begin{figure}[t]
\begin{centering}
\includegraphics[width=\columnwidth]{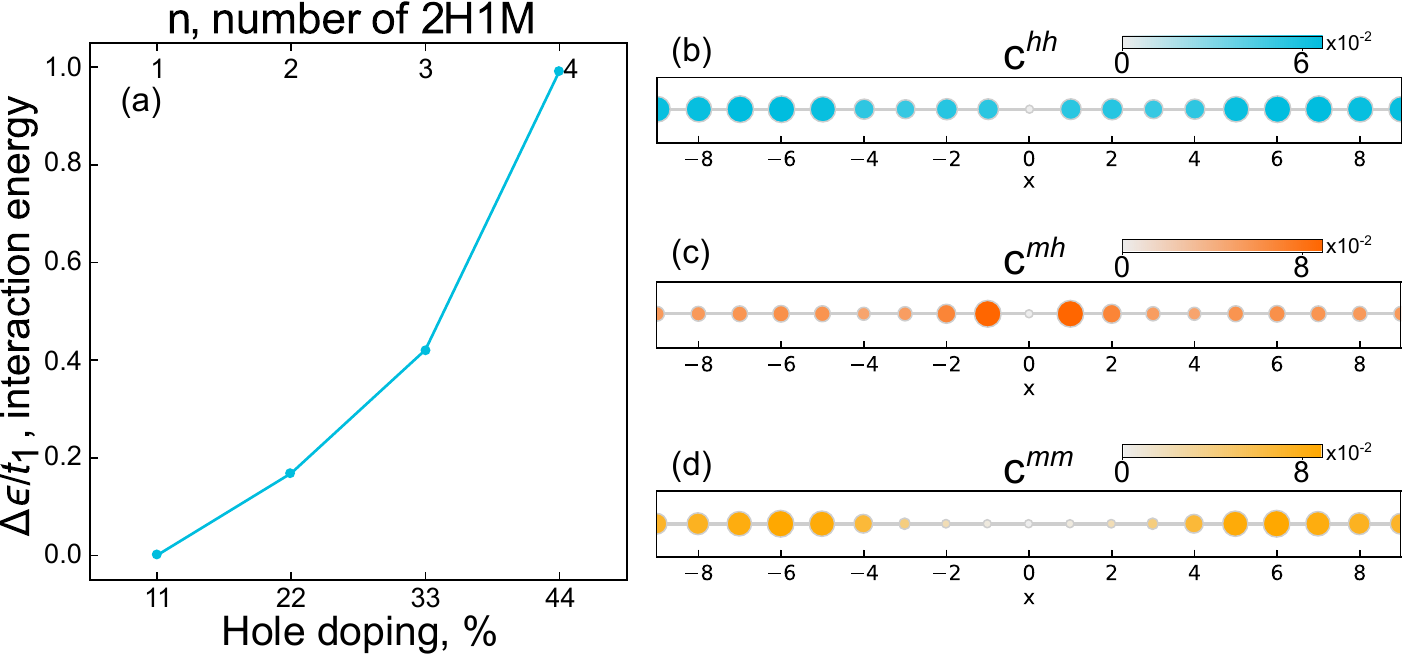}
\par\end{centering}
\caption{ 
\textbf{Magnonic Cooper pairs and their repulsion 0in 1D Hubbard model.} (a) Repulsive interaction energy between the 2H1M particles in 1D Hubbard model with $N=18$ sites and $t_2/t_1 = 0.75, t_3/t_1 = 0.57$. (b) Hole-hole, $c^{hh}$, (c) magnon-hole, $c^{mh}$ and (d) magnon-magnon, $c^{mm}$, correlation functions for three 2H1M particles in 1D. 
}
\label{fig:MCP_interaction_correlation}
\end{figure}

We emphasize that the formation of magnonic Cooper pairs in Hubbard models only relies on the presence of three-fold degenerate band maxima in the energy dispersion. To further illustrate this kinetic mechanism, we investigate another Hubbard model in one dimension. By including up to third neighbor hoppings, we achieve a band structure featuring three local maxima at $\pi/a$ and $\pm k_0$ over a wide range of hopping parameters. By tuning a single parameter, these three band maxima can be made degenerate. Remarkably, our exact diagonalization calculation shows that magnonic Cooper pairs also emerge in this system, precisely when the three band maxima are nearly degenerate. Details are presented in the supplementary materials.

We further study this one-dimensional Hubbard model at finite hole doping density up to $45\%$, with $N=18$ sites. First we look into the interaction energy of $n$ magnonic Cooper pairs, 
\begin{align}
    \Delta\epsilon=E\left(n\times\text{2H1M}\right)-n\cdot E\left(\text{2H1M}\right).
\end{align}
The Fig. \ref{fig:MCP_interaction_correlation}a) shows that the 2H1M particles repel and the interaction energy increases with the number of particles. 
The repulsion is further evidenced by considering the hole-hole, magnon-hole and magnon-magnon correlation functions for three magnonic Cooper pairs as illustrated in the Fig. \ref{fig:MCP_interaction_correlation}b). The presence of repulsive interaction enables magnonic Cooper pairs to Bose condense and form a superconductor at finite doping density.

 We now compare our findings to other mechanisms of strong coupling superconductivity from repulsive interaction. Refs. \cite{Crepel2021, Crepel2022, Crepel2022_2} introduced a mechanism of superconductivity  mediated by virtual excitons in doped insulators. In the narrow band limit, a controlled analysis shows that the pairing energy scale is $t^2/\Delta$, where $\Delta$ represents the insulating gap. 
 In contrast, the pairing energy and $T_{BKT}$ in our case are on the order of the hopping amplitude $t$, which is parametrically larger. 


Similar to the spirit of our work, Ref. \cite{Batista2018} found two-hole-one-magnon bound state in a different triangular lattice Hubbard model with nearest and third-neighbor hoppings. However, in their model this bound state only exists in a narrow range of magnetic field.   
Our work reveals the microscopic origin of the 2H1M bound state formation due to the threefold valley degeneracy of hole dispersion. Moreover, thanks to its large binding energy and stability range, the 2H1M bound state in our model is robust against the introduction of perturbations such as longer range hopping and longer range interaction. This mechanism holds for a wide class systems independent of their details. Additionally, we identify a charge-6$e$ superconductivity from magnonic Cooper pairing.





\begin{acknowledgements}
\textbf{Acknowledgments:} We thank Aidan Reddy and Margarita Davydova for helpful discussions. \textbf{Funding:} This work was supported by National Science Foundation (NSF) Convergence Accelerator Award No. 2235945, and a Simons Investigator Award from the Simons Foundation. L.F. also acknowledges support from the Canadian Institute for Advanced Research (CIFAR). 
\textbf{Author contributions:} K.N. and L.F. contributed essentially to the theoretical analysis and the writing of this work. K.N. performed the ED calculations. \textbf{Competing interests:} The authors declare that they have no competing interests. \textbf{Data and materials availability:} All data needed to evaluate the conclusions in the paper are present in the paper and/or the Supplementary Materials.

\end{acknowledgements}

\bibliography{references}

\begin{widetext}

\newpage

\setcounter{page}{1} 
\appendix

\thispagestyle{empty} 
\begin{center}
    {\Large \textbf{Supplementary Materials for ``Magnonic superconductivity''}}\\[10pt] 
    {\large Kh. G. Nazaryan, L. Fu}\\[5pt] 
    {\large \today}\\[10pt] 
    \vfill
    This file includes additional supportive analyses and figures (S1-S8) which supplement the main findings of the manuscript.
    \vfill
\end{center}
\newpage


\renewcommand{\appendixname}{}

\setcounter{figure}{0}
\renewcommand{\thefigure}{S\arabic{figure}}
\setcounter{section}{0}
\renewcommand{\thesection}{S\arabic{section}}
\renewcommand{\theequation}{S\arabic{equation}}

\begin{center}
    {\large Supporting Material for ``Magnonic superconductivity''}
\end{center}
\begin{center}
    by Kh. G. Nazaryan, L. Fu
\end{center}
\maketitle

\section{Free particle spectrum \label{SecSup:Spectrum}}

In the main text, we discussed the main mechanism that gives an origin to the magnonic Cooper pair formation. In this section, we detail the single particle band structure.

Following our main discourse, we start our from a sufficiently strong magnetic field, which imposes a ferromagnetic ordering into the system. In the case of half-filling, the ferromagnetic ground state is expressed as

\begin{align}
\left|f\right\rangle =\prod_{ \textbf{m} }c_{ \textbf{m} \uparrow}^{\dagger}|0\rangle. 
\end{align}
Here, $|0\rangle$ symbolizes the vacuum state. We have oriented the magnetic field, and consequently, all the spins up.

The system is then doped with a hole or an extra electron. The extra electron will have a spin down and be located at a site already occupied by a spin down electron.

The dispersion of a single hole/electron doped on top of the half-filled system can be easily obtained as
\begin{align}
  &\epsilon(\boldsymbol{\kappa})_{h/e} = h/2 + U\delta_{e} \pm \mu \mp 2 \left( t_1 \epsilon_1(\textbf{k})+t_2 \epsilon_2(\textbf{k})\right),\\
  &\epsilon_1(\textbf{k}) = \cos (k_x a) + \cos \left(k_x a/2 + k_y a\sqrt{3}/2\right) + \cos\left( k_x a/2 - k_y a\sqrt{3}/2\right),\\
   &\epsilon_2(\textbf{k}) = \cos (k_y a \sqrt{3}) + \cos \left(3 k_x a/2 - k_y a\sqrt{3}/2\right) + \cos\left( 3 k_x a/2 + k_y a\sqrt{3}/2\right),
\end{align}

where $\delta_e$ indicates that the $U$ term emerges only for the doped electron (due to double occupancy); $\mu$ is the chemical potential. The electron dispersion is illustrated in the Fig. \ref{figSup:ElDisp}. We infer from this that while the band minimum is always located at $\Gamma$ point, the band maximum is strongly dependent on the parameter $t_2/t_1$. For weak $t_2$, the maximum is double-degenerate and is at $K, K^\prime$, but above a threshold value of $\tau=1/8$, it becomes triple degenerate and moves to $M$ points. 

\begin{figure}[b]
\begin{centering}
\includegraphics[width=0.7\columnwidth]{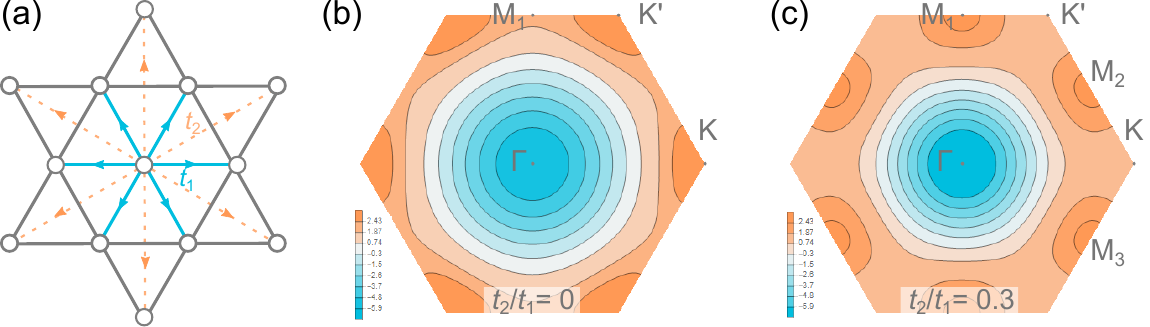}
\par\end{centering}
\caption{\textbf{Modification of electron band structure due to inclusion of second-nearest neighbor hopping.} (a) Schematic  represeantation of the triangular-lattice Hubbard model with the nearest-neighbor, $t_1$, and next-nearest-neighbor, $t_2$, hopping parameters. (b), (c) Are contour plots, illustrating the free electron dispersion within the first BZ for hopping ratios $t_2/t_1=0, 0.3$, respectively. The electron band minimum is at $\Gamma$ point. Meanwhile, the band maximum is double-degenerate at $K, K^\prime$ points for $t_2/t_1 < 1/8$; it becomes triple-degenerate for $t_2/t_1 > 1/8$ shifting to $M$ points.
}
\label{figSup:ElDisp}
\end{figure}

\section{Many-body problem \label{SecSup:Four_Part}}

We are interested in the interplay of two holes doped  into the system. In the context of supercritical magnetic fields, the ground state is expressed as follows,
\begin{align}
    |\Phi ^{2h}\rangle= \sum_{ \textbf{k} , \textbf{l} }\Phi_{\textbf{kl}}|\Phi_{\textbf{kl}}\rangle,\quad|\Phi_{\textbf{kl}}\rangle=c_{ \textbf{k} \uparrow} c_{ \textbf{l} \uparrow}|f\rangle.
\end{align}
 It is an arbitrary linear superposition of states each containing two holes located at cites $\textbf{k}$ and $\textbf{l}$; and the coefficients $\Phi_{\textbf{kl}}$ embody the two-particle wave function.

Upon lowering the magnetic field, in alignment with the principal discourse, the system transitions to another ground state. In this state, one of the spin up electrons flips its spin and hops to another site, creating a hole and a doublon. The wave function corresponding to this excitation is given by,

\begin{align} 
    |\Psi ^{3h1e}\rangle= \sum_{ \textbf{i}, \textbf{j}, \textbf{k}, \textbf{l} }\Psi_{\textbf{ijkl}}|\Psi_{\textbf{ijkl}}\rangle, \quad |\Psi_{\textbf{ijkl}}\rangle=c_{ \textbf{i} \downarrow}^{\dagger}c_{ \textbf{j} \uparrow} c_{ \textbf{k} \uparrow}c_{ \textbf{l} \uparrow} |f\rangle
    \label{eqSup:trimer_wf}
\end{align}

\subsection{\label{SecSup:ShrodEq} Shrodinger's Equation}
\setcounter{equation}{0}





Here we detail the analysis of the 4 particle (1 doublon and 3 holes) state discussed in the main text. 

We act at this state with our Hamiltonian and derive, 
\begin{align} 
    &-t_1 \sum_{ \textbf{i}, \textbf{j}, \textbf{k}, \textbf{l} } \sum_{\boldsymbol{\delta}\in \boldsymbol{\delta}^{(1)}}\Psi_{\textbf{ijkl}}\bigg[|\Psi_{\textbf{i}-\boldsymbol{\delta}\textbf{j}\textbf{k}\textbf{l}}\rangle-|\Psi_{\textbf{i}\textbf{j}+\boldsymbol{\delta}\textbf{k}\textbf{l}}\rangle-|\Psi_{\textbf{i}\textbf{j}\textbf{k}+\boldsymbol{\delta}\textbf{l}}\rangle-|\Psi_{\textbf{i}\textbf{j}\textbf{k}\textbf{l}+\boldsymbol{\delta}}\rangle +\left[\boldsymbol{\delta} \to -\boldsymbol{\delta} \right]\bigg] \nonumber\\ 
    &-t_2\sum_{ \textbf{i}, \textbf{j}, \textbf{k}, \textbf{l} } \sum_{\boldsymbol{\delta}\in \boldsymbol{\delta}^{(2)}}\Psi_{\textbf{ijkl}}\bigg[|\Psi_{\textbf{i}-\boldsymbol{\delta}\textbf{j}\textbf{k}\textbf{l}}\rangle-|\Psi_{\textbf{i}\textbf{j}+\boldsymbol{\delta}\textbf{k}\textbf{l}}\rangle-|\Psi_{\textbf{i}\textbf{j}\textbf{k}+\boldsymbol{\delta}\textbf{l}}\rangle-|\Psi_{\textbf{i}\textbf{j}\textbf{k}\textbf{l}+\boldsymbol{\delta}}\rangle +\left[\boldsymbol{\delta} \to -\boldsymbol{\delta} \right]\bigg]\nonumber\\ 
    &
    +U\sum_{ \textbf{i}, \textbf{j}, \textbf{k}, \textbf{l} } \left(1 - \delta_{\textbf{i}\textbf{j}}-\delta_{\textbf{i}\textbf{k}}-\delta_{\textbf{i}\textbf{l}}\right) \Psi_{\textbf{ijkl}} |\Psi_{\textbf{i}\textbf{j}\textbf{k}\textbf{l}}\rangle
    =\left( E^{3h1e}(t_1, t_2) + 2\mu - 2h 
 \right)\sum_{ \textbf{i}, \textbf{j}, \textbf{k}, \textbf{l} } \Psi_{\textbf{ijkl}} |\Psi_{\textbf{i}\textbf{j}\textbf{k}\textbf{l}}\rangle& \label{eqSup:Shrod_4p}
\end{align}
Here $\boldsymbol{\delta}^{(1)}$  are the three basis vectors on triangular lattice (vectors to nearest neighbors), and $\boldsymbol{\delta}^{(2)}$ are the vectors to next-nearest neighbors.


We note that the Pauli exclusion principle leads the antisymmetry of the wave function under the swap of two holes,
\begin{align} 
    |\Psi_{\textbf{ijkl}}\rangle = - |\Psi_{\textbf{ikjl}}\rangle = -|\Psi_{\textbf{ijlk}}\rangle 
\end{align}
We introduce a linear coordinate that encompasses the position of each particle on the two-dimensional lattice,
\begin{align} 
    \alpha_m = L (x_m -1) + (y_m -1) +1, \quad \alpha_m \in [1,L^2], \quad m=i,j,k,l. 
\end{align}
The initial 2D coordinates of the given particles are then identically recovered from these linear coordinates. 
Then we can generate basis states for which the linear coordinate of the hole $\textbf{j}$ is smaller than that of hole $\textbf{k}$, $\alpha_j < \alpha_k$, and similarly, $\alpha_k<\alpha_l$,
\begin{align}
    \alpha_j<\alpha_k<\alpha_l
\end{align}
By Pauli exclusion principle, these states span the entire Hilbert space.
 
We then proceed by projecting the Eq. \eqref{eqSup:Shrod_4p} onto the bra- state to obtain a set of equations on the coefficients $\Psi_\textbf{ijkl}$: 
\begin{align} 
    &-t_1\sum_{\boldsymbol{\delta}\in \boldsymbol{\delta}^{(1)}}\bigg[\left(\Psi_{\textbf{i}+\boldsymbol{\delta}\textbf{j}\textbf{k}\textbf{l}}-\Psi_{\textbf{i}\textbf{j}-\boldsymbol{\delta}\textbf{k}\textbf{l}} -\Psi_{\textbf{i}\textbf{j}\textbf{k}-\boldsymbol{\delta}\textbf{l}}-\Psi_{\textbf{i}\textbf{j}\textbf{k}\textbf{l}-\boldsymbol{\delta}}\right) +\left[\boldsymbol{\delta} \to -\boldsymbol{\delta} \right]\bigg]\nonumber\\
    &-t_2\sum_{\boldsymbol{\delta}\in \boldsymbol{\delta}^{(2)}}\bigg[\left(\Psi_{\textbf{i}+\boldsymbol{\delta}\textbf{j}\textbf{k}\textbf{l}}-\Psi_{\textbf{i}\textbf{j}-\boldsymbol{\delta}\textbf{k}\textbf{l}} -\Psi_{\textbf{i}\textbf{j}\textbf{k}-\boldsymbol{\delta}\textbf{l}} -\Psi_{\textbf{i}\textbf{j}\textbf{k}\textbf{l}-\boldsymbol{\delta}}\right) +\left[\boldsymbol{\delta} \to -\boldsymbol{\delta} \right]\bigg]\nonumber\\
    &+
    U\sum_{ \textbf{i}, \textbf{j}, \textbf{k}, \textbf{l} } \left(1 - \delta_{\textbf{i}\textbf{j}}-\delta_{\textbf{i}\textbf{k}}-\delta_{\textbf{i}\textbf{l}}\right) \Psi_{\textbf{ijkl}}
    =\left( E^{3h1e}(t_1, t_2) + 2\mu - 2h 
 \right) \Psi_{\textbf{ijkl}} 
\end{align}

In the limit of infinite $U$, the double occupancy is forbidden. And thus, the doublon must be located at the same site with one of the holes (spin flip channel). In other words, the condition for one of the Kronecker deltas must always be satisfied,
\begin{align}
    \alpha_i = \alpha_j, \quad \text{or} \quad \alpha_i = \alpha_k, \quad \text{or} \quad \alpha_i = \alpha_l.
\end{align}
This substantially reduces the Hilbert space dimension, allowing to simulate larger system sizes.

We also note that a similar approach will yield equations for any number of holes and doublons. The resulting equations can be easily recovered in the general case by noticing that the doublons hop with a coefficient $-t_1$ and the holes with $+t_1$.  


\subsection{Fixed Center of Mass Momentum}

In the discussed problem the center of mass (COM) momentum serves as a good quantum number. To exploit this, we construct the basis of Fock states in the site occupation representation. Then construct a COM translation operator $T_\textbf{R}$ which shifts all the particles in the system by $\textbf{R}$, and thus preserves the relative coordinates (mod s). For every Fock state we apply $T_\textbf{R}$ to see what states it is related to and sort all Fock states into cycles. 
The eigenstates of the translation operator are then constructed from each of these cycles as  
\begin{align}
    \Psi^{n}_{\textbf{P}}(\textbf{r}_{ij},\textbf{r}_{ik},\textbf{r}_{il})=\frac{1}{N_{n}}\sum_{\textbf{R}}e^{-i\textbf{P}\textbf{R}}T_\textbf{R}\Psi(0,\textbf{r}_{ij},\textbf{r}_{ik},\textbf{r}_{il}),
\end{align}
where $N_{n}$ is the length of the cycle $n$. Each cycle has $N_{n}$ allowed values for $\textbf{P}$ which are determined from the periodicities under applying the translation operator $T_\textbf{R}$.

We then decompose the wave function into a superposition of these states and apply the Hamiltonian to construct the corresponding matrix and determine the spectrum. 
This approach is applicable in general case but often is computationally intensive. 

In the case when we have a single doublon, we can employ a more optimized version of this approach by fixing the COM coordinate with the doublon and introducing the relative ones as,
\begin{align} 
    \textbf{R}= \textbf{i}, \quad \textbf{r}_{ij} = \textbf{j}-\textbf{i}, \quad \textbf{r}_{ik} = \textbf{k}-\textbf{i}, \quad \textbf{r}_{il} = \textbf{l}-\textbf{i}.
\end{align}

We carry out a Fourier transform over $\textbf{R}$, 
\begin{align}
    \Psi_{\textbf{P}}(\textbf{r}_{ij},\textbf{r}_{ik},\textbf{r}_{il})=\frac{1}{\sqrt{N}}\sum_{\textbf{R}}e^{-i\textbf{P}\textbf{R}}\Psi(\textbf{R},\textbf{r}_{ij},\textbf{r}_{ik},\textbf{r}_{il}),  
\end{align}
to obtain equations for a fixed center of mass momentum $\textbf{P}$. They can be written as,

\begin{align} 
    &-t_1\sum_{\boldsymbol{\delta}\in \boldsymbol{\delta}^{(1)}}\bigg[\left(e^{i\textbf{P}\boldsymbol{\delta}}\Psi_{\textbf{P}}\left(\textbf{r}_{ij}-\boldsymbol{\delta},\textbf{r}_{ik}-\boldsymbol{\delta},\textbf{r}_{il}-\boldsymbol{\delta}\right)-\Psi_{\textbf{P}}\left(\textbf{r}_{ij}+\boldsymbol{\delta},\textbf{r}_{ik},\textbf{r}_{il}\right) -\Psi_{\textbf{P}}\left(\textbf{r}_{ij},\textbf{r}_{ik}+\boldsymbol{\delta},\textbf{r}_{il}\right)
    -\Psi_{\textbf{P}}\left(\textbf{r}_{ij},\textbf{r}_{ik},\textbf{r}_{il}+\boldsymbol{\delta}\right)\right)+\text{h.c.}\bigg]&\nonumber\\
    &-t_2\sum_{\boldsymbol{\delta}\in \boldsymbol{\delta}^{(2)}}\bigg[\left(e^{i\textbf{P}\boldsymbol{\delta}}\Psi_{\textbf{P}}\left(\textbf{r}_{ij}-\boldsymbol{\delta},\textbf{r}_{ik}-\boldsymbol{\delta},\textbf{r}_{il}-\boldsymbol{\delta}\right)-\Psi_{\textbf{P}}\left(\textbf{r}_{ij}+\boldsymbol{\delta},\textbf{r}_{ik},\textbf{r}_{il}\right) -\Psi_{\textbf{P}}\left(\textbf{r}_{ij},\textbf{r}_{ik}+\boldsymbol{\delta},\textbf{r}_{il}\right)
    -\Psi_{\textbf{P}}\left(\textbf{r}_{ij},\textbf{r}_{ik},\textbf{r}_{il}+\boldsymbol{\delta}\right)\right)+\text{h.c.}\bigg]&\nonumber\\
    &\quad + U\left(1 - \delta_{\textbf{r}_{ij}0}-\delta_{\textbf{r}_{ik}0}-\delta_{\textbf{r}_{il}0}
\right)\Psi_{\textbf{P}}\left(\textbf{r}_{ij},\textbf{r}_{ik},\textbf{r}_{il}\right) = \left( E^{3h1e}(t_1, t_2) + 2\mu - 2h 
 \right) \Psi_{\textbf{P}}\left(\textbf{r}_{ij},\textbf{r}_{ik},\textbf{r}_{il}\right)& \label{eqSup:4Part}
\end{align}

We can again introduce the linear variables $\alpha_{ij}, \alpha_{ik}, \alpha_{il}$. The condition to cover all the Hilbert space is
\begin{align}
    \alpha_{ij}<\alpha_{ik}<\alpha_{il}
\end{align}
We note that because of this condition the second and third Kronecker deltas in the above equation will always be 0. 

The fixed CM momentum becomes even more useful when dealing with the infinite $U$ limit. 
The corresponding equation is can be obtained in a similar manner by considering a wave function with a spin flip and two holes,
\begin{align}
    &|\psi\rangle=\sum_{ijk}\psi_{i,jk}S_{i}^{\dagger}c_{j\uparrow}c_{k\uparrow}\left|f\right\rangle 
\end{align}
Here the $S_i^\dagger=c_{i\downarrow}^{\dagger}c_{i\uparrow}$ is spin flip creation operator at site $i$.

Acting by the Hamiltonian we obtain equations for the coefficients,
\begin{align}
    &t_1\sum_{\boldsymbol{\delta}\in \boldsymbol{\delta}^{(1)}}\left(\psi_{\textbf{i},\textbf{j}+\delta \textbf{k}} +\psi_{ \textbf{i},\textbf{jk}+\delta}\right)
    + t_2\sum_{\boldsymbol{\delta}\in \boldsymbol{\delta}^{(2)}} \left(\psi_{\textbf{i},\textbf{j}+\delta \textbf{k}}
    +\psi_{ \textbf{i},\textbf{jk}+\delta}\right)\nonumber\\
    &+t_1\sum_{\boldsymbol{\delta}\in \boldsymbol{\delta}^{(1)}}\left(\psi_{\textbf{j},\textbf{ik}}\delta\left(\textbf{i}-\textbf{j}-\delta\right)+\psi_{\textbf{k},\textbf{ji}}\delta\left(\textbf{i}-\textbf{j}-\delta\right)\right) + t_2\sum_{\boldsymbol{\delta}\in \boldsymbol{\delta}^{(2)}}\left(\psi_{\textbf{j},\textbf{ik}}\delta\left(\textbf{i}-\textbf{j}-\delta\right)+\psi_{\textbf{k},\textbf{ji}}\delta\left(\textbf{i}-\textbf{j}-\delta\right)\right) \nonumber\\
    &=\left( E^{3h1e}(t_1, t_2) + 2\mu - 2h 
 \right)\psi_{\textbf{i},\textbf{jk}}
\end{align}
Here the first line describes the hole hopping, while the second line stands for the spin flip hopping to a hole if they are on neighboring sites.  

We then fix the COM coordinate with the spin flip
\begin{align}
    \psi_{\textbf{i},\textbf{jk}}=\psi\left(\textbf{R};\textbf{R}+\textbf{r}_{ij},\textbf{R}+\textbf{r}_{ik}\right),\quad \textbf{R}=\textbf{i},\quad \textbf{r}_{ij}=\textbf{j}-\textbf{i},\quad \textbf{r}_{ik}=\textbf{k}-\textbf{i}
\end{align}

For the COM momentum representation we carry out a Fourier transform
\begin{align}
\psi\left(\textbf{R};\textbf{R}+\textbf{r}_{ij},\textbf{R}+\textbf{r}_{ik}\right) =\sum_{\textbf{P}}e^{-i\textbf{PR}}\psi_{\textbf{P}}\left(\textbf{r}_{ij},\textbf{r}_{ik}\right)
\end{align}
the equation then reads as 
\begin{align}
   &\left(E^{3h1e}(t_1,t_2)+2\mu - 2h\right)\psi_{\textbf{P}}\left(\textbf{r}_{ij},\textbf{r}_{ik}\right)- \nonumber\\
    &-t_1\sum_{\boldsymbol{\delta}\in \boldsymbol{\delta}^{(1)}}\left(\psi_{\textbf{P}}\left(\textbf{r}_{ij}+\delta,\textbf{r}_{ik}\right)+\psi_{\textbf{P}}\left(\textbf{r}_{ij},\textbf{r}_{ik}+\delta\right)\right)
    -t_2\sum_{\boldsymbol{\delta}\in \boldsymbol{\delta}^{(2)}}\left(\psi_{\textbf{P}}\left(\textbf{r}_{ij}+\delta,\textbf{r}_{ik}\right)+\psi_{\textbf{P}}\left(\textbf{r}_{ij},\textbf{r}_{ik}+\delta\right)\right)=
    \\
   &=t_1\sum_{\boldsymbol{\delta}\in \boldsymbol{\delta}^{(1)}}\left(\psi_{\textbf{P}}\left(-\textbf{r}_{ij},-\textbf{r}_{ij}+\textbf{r}_{ik}\right)\delta\left(\textbf{r}_{ij}-\delta\right)e^{i\boldsymbol{P\delta}}+\psi_{\textbf{P}}\left(-\textbf{r}_{ik}+\textbf{r}_{ij},-\textbf{r}_{ik}\right)\delta\left(\textbf{r}_{ik}-\delta\right)e^{i\boldsymbol{P\delta}}\right)\nonumber\\
   & +t_2\sum_{\boldsymbol{\delta}\in \boldsymbol{\delta}^{(2}}\left(\psi_{\textbf{P}}\left(-\textbf{r}_{ij},-\textbf{r}_{ij}+\textbf{r}_{ik}\right)\delta\left(\textbf{r}_{ij}-\delta\right)e^{i\boldsymbol{P\delta}}+\psi_{\textbf{P}}\left(-\textbf{r}_{ik}+\textbf{r}_{ij},-\textbf{r}_{ik}\right)\delta\left(\textbf{r}_{ik}-\delta\right)e^{i\boldsymbol{P\delta}}\right)
\end{align}

\subsection{Exact Diagonalization (ED)}

\setcounter{equation}{0}

To employ ED we construct the hopping matrices described above. Though their linear size increases rapidly with the system size, the matrices remain very sparse. This allows us to employ algorithms specifically designed for storing (compressed sparse row CSR) and manipulating large sparse matrices. Namely, the first several eigenvalues and eigenvectors are found by employing Lanczos algorithm. 

We implemented our code both in Wolfram Mathematica and Python to assure an exact agreement of the numerical results. Additionally, we developed two different methods for ED in real space, a general code which deals with any given number of holes and doublons, as well as specific algorithm which works more efficiently for the 3 hole 1 doublon problem, allowing to scale the system size. They both agreed exactly on every simulated scenario.

The both algorithms were implemented for finite and infinite $U$ cases. We made sure that the results in the large $U$ limit converged to the infinite $U$ results seamlessly. For infinite $U$ limit, we were able to scale the system size up to $N=18\times 18$. 

Separately, we implemented the ED code in the basis of fixed CM momentum as well. Thanks to the reduced Hilbert space dimension, we were able to reach a rather large system size of $N=36\times 36$ for the 2H1S problem. We checked that the results for smaller system sizes agree exactly with the direct ED calculations.

\section{Robustness of results}

In this section we examine the 2H1M phase, discussed in the main text, and show that it is a robust property of the our model, resilient within infinite lattice systems.
To conclusively establish this we explore the energy gap observed in the Fig. \ref{fig:Band_Structure}a),b) of the main text. We analyze the width of the gap as a function of the system size and extend of our findings from finite lattices to infinite lattice configurations.  The extrapolated data corresponding to finite systems is graphically represented in Fig. \ref{figSup:FiniteSize}a). For small enough $t_2/t_1$, when the band minimum is at $K$ point, the corresponding gap shrinks as $\Delta_K \propto L^{-2}$, eventually vanishing for an infinite system. Conversely, the gap width demonstrates a different pattern for larger $t_2$, when the band minima shift to $M$ or $\Gamma$ points. While here the gap again follows a $L^{-2}$ scaling with a reasonable accuracy, it saturates to a finite value in the infinite size limit, 
\begin{align}
    & \Delta_M \approx 0.146 t_1, \quad t_2 = 0.09 t_1, \\
    & \Delta_\Gamma \approx 0.375 t_1, \quad t_2 = 0.15 t_1.
\end{align}
This delivers compelling evidence that this energy gain is a phenomenon inherent to infinitely large systems as well, thus affirming the robustness of the predicted phase as an intrinsic  feature of the Hubbard model.

Additionally, we illustrate the dependency of the average distance between the holes on the parameter $t_2/t_1$ in the Fig. \ref{figSup:FiniteSize}b), and compare the results for systems of increasing linear size $L$. 
we observe that, initially, at small $t_2$, the holes are far apart (of order of system size), indicating that one of the holes bounds to the spin flip, forming a polaron, while the other hole is repelled from that pair. Here the average size scales with the system size.

When $t_2/t_1$ exceeds $0.09$, the localization length of the ground state abruptly drops, reaching $|\Delta r|/a \sim 3$. This occurs because the GS transitions to the $P=M$ sector, which was already localized even for smaller $t_2$. Importantly, here the pair size shows very little sensitivity to the system size, proving the state to be well-localized.

\begin{figure}[h!]
\begin{centering}
\includegraphics[width=0.9\columnwidth]{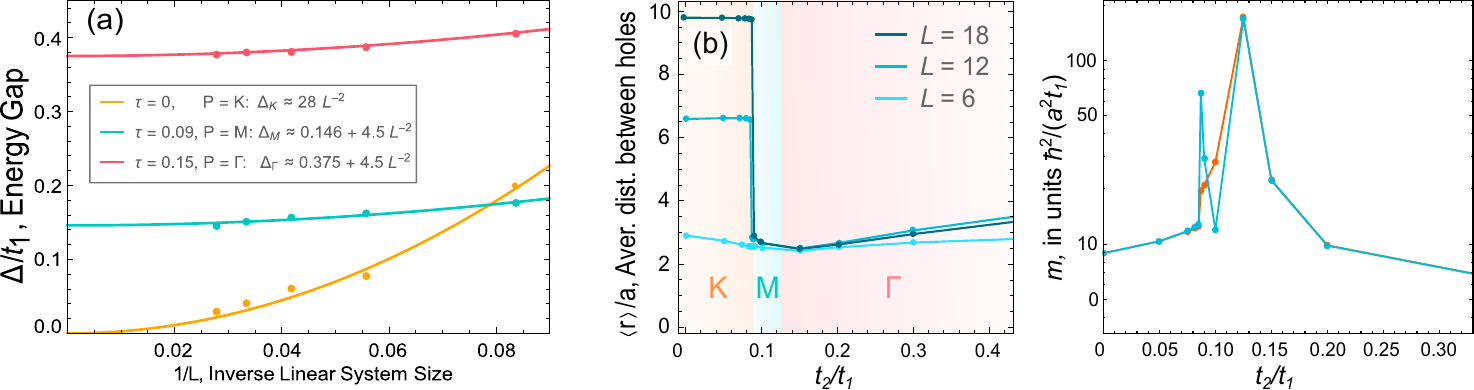}
\par\end{centering}
\caption{\textbf{Properties of magnonic Cooper pair particles.} (a) Explores the energy gaps at fixed CM momentum as a function of inverse linear system size $1/L$, when the corresponding state is the GS. The extrapolation reveal that, for $t_2 = 0$ the gap vanishes for infinite systems, while for $t_2/t_1 = 0.09, 0.15$ saturates to finite values. (b) The average distance between the holes as a function of $t_2/t_1$ for increasing system size. The state becomes well-localized for $t_2/t_1\gtrsim 0.09$, and the GS momentum shifts from $K$ to $M$, then t $\Gamma$. (c) Effective masses of the 2H1M particle $m_x, m_y$. The masses are anisotropic when the GS is at $M$ points. 
}
\label{figSup:FiniteSize}
\end{figure}

\section{Finite on-site interaction $U$} 

Our main text focused on the limit of infinite on-site interaction energy $U/t_1$. As discussed above, the Hilbert space dimension is substantially reduced in that case, allowing for more reliable calculations for larger system sizes. 
In this section, we present the results for finite $U/t_1$ case. From our main discussion we know that the $K$ and $M$ points in momentum space play a pivotal role in our system. Thus when fixing the finite system size $L$, we need to choose it in a way that both $K$ and $M$ points are reachable, i.e. $L$ should be divisible by $6$. And since $12\times 12$ is very computationally intensive, the largest system size that we can have is $N=6\times 6$.

We illustrate that the remarkable property of 2H1M bound state manifests in the case of finite interaction $U$ as well. In the Fig. \ref{figSup:FiniteU} we illustrate the energy gap between the 2H1M state and the (1H1M + 1H) state (polaron and hole):
\begin{align}
    \Delta = \epsilon^{2H1M} - (\epsilon^{pol} + \epsilon^{H}) \label{eqSup:TrimVS(PolHole)}.
\end{align}
We observe that for any interaction $U$, the localized magnonic Cooper pair state becomes more favorable for large enough second-neighbor hopping parameter. Interestingly, while for large enough $U$ the localization takes place near $t_2/t_1 \approx 0.09$, in agreement with the infinite $U$ results, by lowering the $U$, the threshold value is gradually pushed towards exactly $t_2 = t_1/8$. This result is in a perfect agreement with the intuitive picture of the pairing mechanism involving the holes shifting to the $M$ points, described in the main text. 
Additionally, we see that the gap has a local maximum for intermediate interactions of $U/t_1 \sim 15$.


\begin{figure}[h!]
\begin{centering}
\includegraphics[width=0.3\columnwidth]{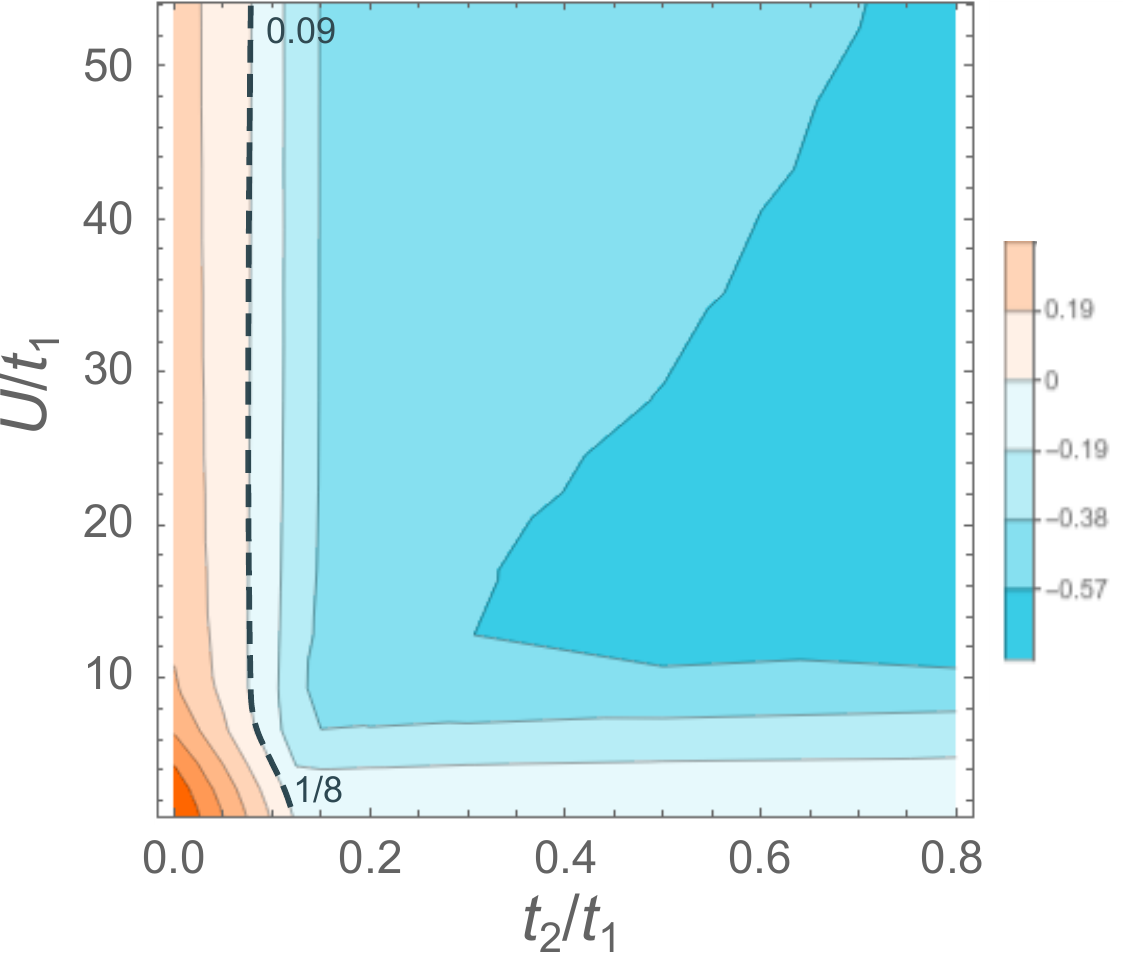}
\par\end{centering}
\caption{\textbf{Binding energy of the Magnonic Cooper pairs for finite $U$.} Energy gap between the 2H1M and the polaron and hole (1H1M + 1H) state, Eq. \eqref{eqSup:TrimVS(PolHole)} as a function of $U/t_1$ and $t_2/t_1$. Dashed line shows the transition to bound state.
}
\label{figSup:FiniteU}
\end{figure}


\section{Interactions between 2H1M particles} 

 As discussed in the main text, we determined the sign of interactions of the magnonic Cooper pairs by performing an ED study for a system of 4 holes and 2 magnons on a $6\times 6$ lattice. We showed that the net interaction is attractive for these particles. 

Here we present the analysis for the effective model described in the main text for studying the interaction of effective bosons. 
To illustrate of the localization of the bosons into a composite particles in the low doping regime, we consider a triangular lattice model with nearest neighbor hopping. In the long-wavelength regime the internal structure of the lattice is rendered unimportant, but the choice of the triangular lattice facilitates the modeling of the anisotropic spectrum of the 2H1M bosons. We obtained that these bosons have anisotropic masses, typically with $m_y/m_x \sim 2$. We orient eigen-axis for the type 1 boson dispersion, $\alpha=1$, along the $x$ and $y$ axis with masses $m_x$ and $m_y$. For the boson 2 and 3 we can then rotate these axes by $\pi/3$. In other words, for the boson of type 1, as an example, we model hopping with parameter $t$ in the $x$ direction, and $t^\prime$ in the two other directions. In the long wavelength limit, the spectrum is expanded as
\begin{align}
    \epsilon_{b,1} = \frac{p_x^2}{2m_x} + \frac{p_y^2}{2m_y}, \quad m_x = \frac{\hbar^2}{a^2 (2 t + t^\prime)},\quad m_y = \frac{\hbar^2}{3 a^2 t^\prime} 
\end{align}
To obtain $m_y/m_x \sim 2$ for 2H1M boson, we choose $t^\prime /t =1/2$. 
The anisotropic spectra for the other two type bosons are modeled in a similar manner. 

In order to show the localization, it is enough to perform ED analysis with at most a single boson of each type. Thus, in the Hamiltonian from the Eq.\eqref{eq:Effective_FT} of the main text, the $U, J$ terms are left out as they involve interactions of two bosons of the same type. We model the attractive $V$ term on the lattice as on-site interaction with magnitude $V_0$. 

\subsection{Two boson localization}

We start from the two boson localization on the example of $\alpha = 1,2$ type bosons.  
Similarly to the analysis in the Sec. \ref{SecSup:ShrodEq}, we can write the Shr\"odinger equation for the two particle wave function $\Phi_\textbf{ij}$  in the form
\begin{align}
&-t\left(\Phi_{\left(\textbf{i}+\boldsymbol{\delta}_{1}\right)\textbf{j}}+\Phi_{\left(\textbf{i}-\boldsymbol{\delta}_{1}\right)\textbf{j}}\right)-t^{\prime}\sum_{m=2,3}\left(\Phi_{\left(\textbf{i}+\boldsymbol{\delta}_{m}\right)\textbf{j}}+\Phi_{\left(\textbf{i}-\boldsymbol{\delta}_{m}\right)\textbf{j}}\right)\\
&\quad -t\left(\Phi_{\textbf{i}\left(\textbf{j}+\boldsymbol{\delta}_{2}\right)}+\Phi_{\textbf{i}\left(\textbf{j}-\boldsymbol{\delta}_{2}\right)}\right)-t^{\prime}\sum_{m=1,3}\left(\Phi_{\textbf{i}\left(\textbf{j}+\boldsymbol{\delta}_{m}\right)}+\Phi_{\textbf{i}\left(\textbf{j}-\boldsymbol{\delta}_{m}\right)}\right)-V_{0}\Phi_{\textbf{ij}}\delta_{\textbf{ij}}=E^{(2)}\Phi_{\textbf{ij}}
\end{align}

The two-particle problem can be conveniently studied by changing the initial variables to the center of mass $\textbf{R}$ and relative motion $\textbf{l}$, 
\begin{align}  
    \Phi_{\textbf{ij}}=\Phi_\textbf{Rl},\quad \textbf{R}=\textbf{i},\quad \textbf{l}=\textbf{j}-\textbf{i}
\end{align}
This will rewrite the equation above in the form
\begin{align}
    &-t\left(\Phi_{\left(\textbf{R}+\boldsymbol{\delta}_{1}\right)\left(\textbf{l}-\boldsymbol{\delta}_{1}\right)}+\Phi_{\left(\textbf{R}-\boldsymbol{\delta}_{1}\right)\left(\textbf{l}+\boldsymbol{\delta}_{1}\right)}\right)-t^{\prime}\sum_{m=2,3}\left(\Phi_{\left(\textbf{R}+\boldsymbol{\delta}_{m}\right)\left(\textbf{l}-\boldsymbol{\delta}_{m}\right)}+\Phi_{\left(\textbf{R}-\boldsymbol{\delta}_{m}\right)\left(\textbf{l}+\boldsymbol{\delta}_{m}\right)}\right)\nonumber\\&\quad -t\left(\Phi_{\textbf{R}\left(\textbf{l}+\boldsymbol{\delta}_{2}\right)}+\Phi_{\textbf{R}\left(\textbf{l}-\boldsymbol{\delta}_{2}\right)}\right)-t^{\prime}\sum_{m=1,3}\left(\Phi_{\textbf{R}\left(\textbf{l}+\boldsymbol{\delta}_{m}\right)}+\Phi_{\textbf{R}\left(\textbf{l}-\boldsymbol{\delta}_{m}\right)}\right)-V_{0}\Phi_{\textbf{Rl}}\delta_{\textbf{l}}=E^{(2)}\Phi_{\textbf{Rl}}
\end{align}

We can then carry out a Fourier transform over $\textbf{R}$, $\Phi_{\textbf{P}}(\textbf{l})=\frac{1}{\sqrt{N}}\sum_{\textbf{R}}e^{-i\textbf{P}\textbf{R}}\Phi_{\textbf{R}\textbf{l}}$, to obtain equations for a fixed center of mass momentum $\textbf{P}$.
The resulting equations read as follows:
\begin{align}
    &-\left(te^{i\textbf{P}\boldsymbol{\delta}_{1}}+t^{\prime}\right)\Phi_{\textbf{P}}\left(\textbf{l}-\boldsymbol{\delta}_{1}\right)-\left(te^{-i\textbf{P}\boldsymbol{\delta}_{1}}+t^{\prime}\right)\Phi_{\textbf{P}}\left(\textbf{l}+\boldsymbol{\delta}_{1}\right)\nonumber\\
    &\quad -\left(t^{\prime}e^{i\textbf{P}\boldsymbol{\delta}_{2}}+t\right)\Phi_{\textbf{P}}\left(\textbf{l}-\boldsymbol{\delta}_{2}\right)-\left(t^{\prime}e^{-i\textbf{P}\boldsymbol{\delta}_{2}}+t\right)\Phi_{\textbf{P}}\left(\textbf{l}+\boldsymbol{\delta}_{2}\right)\nonumber\\
    &\quad -t^{\prime}\left(e^{i\textbf{P}\boldsymbol{\delta}_{3}}+1\right)\Phi_{\textbf{P}}\left(\textbf{l}-\boldsymbol{\delta}_{3}\right)-t^{\prime}\left(e^{-i\textbf{P}\boldsymbol{\delta}_{3}}+1\right)\Phi_{\textbf{P}}\left(\textbf{l}+\boldsymbol{\delta}_{3}\right) -V_{0}\Phi_{\textbf{P}}\left(\textbf{l}\right)\delta_{\textbf{l}}=E\Phi_{\textbf{P}}({\textbf{l}})
\end{align}

This equation has two different families of solution, depending on whether $\Phi_\textbf{P}(\textbf{l}=0)$ is zero or finite. In the first case, there is no overlap between the wave functions of the two bosons, while in the second case the bosons are bound. Here we will analyze them and quantify the transition from one to another depending on the parameters of the model. 
 
(1) For the first case, $\Phi_\textbf{P}(\textbf{l}=0)=0$, we can proceed easily by carrying out another Fourier transform over the relative coordinate $\textbf{l}$, as $\tilde{\Phi}_{\textbf{P}}(\textbf{k})=\frac{1}{\sqrt{N}}\sum_{\textbf{l}}e^{-i\textbf{k}\textbf{l}}\Phi_{\textbf{P}}(\textbf{l})$. Then the equations are algebraic and can be solved to yield the dispersion for this case:
\begin{align} 
E^{(2)}_\text{unb}(\textbf{P},\textbf{k})&=\left(te^{i\textbf{P}\boldsymbol{\delta}_{1}}+t^{\prime}\right)e^{-i\textbf{k}\boldsymbol{\delta}_{1}}+\left(te^{-i\textbf{P}\boldsymbol{\delta}_{1}}+t^{\prime}\right)e^{i\textbf{k}\boldsymbol{\delta}_{1}}\nonumber\\
&+\left(t^{\prime}e^{i\textbf{P}\boldsymbol{\delta}_{2}}+t\right)e^{-i\textbf{k}\boldsymbol{\delta}_{2}}+\left(t^{\prime}e^{-i\textbf{P}\boldsymbol{\delta}_{2}}+t\right)e^{i\textbf{k}\boldsymbol{\delta}_{2}}\nonumber\\
&+t^{\prime}\left(e^{i\textbf{P}\boldsymbol{\delta}_{3}}+1\right)e^{-i\textbf{k}\boldsymbol{\delta}_{3}}+t^{\prime}\left(e^{-i\textbf{P}\boldsymbol{\delta}_{3}}+1\right)e^{i\textbf{k}\boldsymbol{\delta}_{3}}
\end{align}
In this case the bosons move absolutely independently. It is easy to show that the minimum energy is achieved for COM momentum $\textbf{P}=0$, and the long-wavelength expansion for the dispersion reads as
\begin{align} 
&E^{(2)}_\text{unb} (\textbf{k})\approx -4(t+2t^\prime) + \frac{\hbar^2 q_x^2}{2M_1}+\frac{\hbar^2 q_y^2}{2M_2}, \quad M_1=\frac{\hbar^2}{3 a^2(t+t^\prime)}; \quad M_2=\frac{\hbar^2}{a^2(t+5 t^\prime)}, \\
& \begin{pmatrix}q_{x}\\
q_{y}
\end{pmatrix}=\begin{pmatrix}\cos\phi & -\sin\phi\\
\sin\phi & \cos\phi
\end{pmatrix}\begin{pmatrix}k_{x}\\
k_{y}
\end{pmatrix}, \quad \phi =-\pi/6\nonumber. 
\end{align}
With the minimal energy being $E^{(2)}_\text{unb,min} =-4(t+2t^\prime)$. We note that this is simply twice the GS energy of a single boson $E^{(1)}_\text{min} =-2(t+2t^\prime)$.

(2) In the case when $\Phi_\textbf{P}(\textbf{l}=0)\neq 0$, we again carry out a Fourier transform. Here we have an additional term arising from $\delta_{ \textbf{l} 0}$. It can be Fourier transformed as, $\frac{1}{\sqrt{N}}\sum_{\textbf{l}}\delta_{\textbf{l}0}f(\textbf{l})e^{-i\textbf{k}\textbf{l}}=\frac{1}{\sqrt{N}}f(\textbf{l}=0)$. 
The resulting equation is then written as follows,
\begin{align} \refstepcounter{equation}\tag{A3.\arabic{equation}}
    &E^{(2)}_\text{b}  \tilde{\Phi}_{\textbf{P}}(\textbf{k})=E^{(2)}_\text{unb}(\textbf{P},\textbf{k}) \tilde{\Phi}_{\textbf{P}}(\textbf{k})-\frac{V_0}{\sqrt{N}}\Phi_{\textbf{P}}(\textbf{l}=0)
\end{align}
We then express,
\begin{align}
    & \tilde{\Phi}_{\textbf{P}}(\textbf{k}) = - \frac{V_0}{\sqrt{N}}\frac{\Phi_{\textbf{P}}(\textbf{l}=0)}{E^{(2)}_\text{b}  - E^{(2)}_\text{unb}(\textbf{P},\textbf{k})}
\end{align}
Carrying out an inverse Fourier transform we obtain,
\begin{align}
    \Phi_{\textbf{P}}(\textbf{l})=\frac{1}{\sqrt{N}}\sum_{\textbf{k}}e^{i \textbf{k} \textbf{l} }\tilde{\Phi}_{\textbf{P}}(\textbf{k}) = 
    - \frac{V_0}{N} \Phi_{\textbf{P}}(\textbf{l}=0) \sum_{\textbf{k}}\frac{e^{i \textbf{k} \textbf{l} }}{E^{(2)}_\text{b} - E^{(2)}_\text{unb}(\textbf{P},\textbf{k})}
\end{align}
Now we fix $\textbf{l} =0$ to obtain a self-consistency equation on the energy $E^{(2)}_\text{b}$,
\begin{align}
    1= 
    - \frac{V_0}{N} \sum_{\textbf{k}}\frac{1}{E^{(2)}_\text{b}  - E^{(2)}_\text{unb}(\textbf{P},\textbf{k})}
    \label{eqSup:Second_type}
\end{align}

Depending on the parameters of the system, either of the two above solutions can have lower energy. Let us find the critical value of the on-site interaction term $V^\text{c}_{0}$, after which the bound solution is more favorable than the unbound one. 

The lowest energy from the unbound solution is, as we found, $E^{(2)}_\text{unb, min} = -4(t+2t^\prime)$. At $V_0=V^\text{c}_{0}$, it should coincide with the energy of bound solution. So, we plug it into the Eq. \eqref{eqSup:Second_type}, evaluate the sum and express $V^\text{c}_{0}$. 

The summation can be replaced by integration as $\sum_{\textbf{k}}\to\frac{A}{(2\pi)^{2}}\int_{BZ}d^{2} \textbf{k}$, where $A$ is the sample surface area. Additionally, since the second term in the denominator Eq. has a minimum at $ \textbf{k} = 0$, we expand the denominator up to the second order to arrive to the following integral,
\begin{align} 
    1\approx \frac{V^\text{c}_0}{N}\frac{A}{(2\pi)^{2}}\int_{BZ}d^{2}  \textbf{q} \frac{1}{\frac{\hbar^2 q_x^2}{2M_1}+\frac{\hbar^2 q_y^2}{2M_2}}
\end{align}

We notice that the last integral diverge logarithmically at small $q$. To avoid the divergence, we introduce a cutoff at the inverse lattice size $q\sim 1/(L a)$. Within the logarithmic precision, the integral is evaluated as,
\begin{align} 
    1\approx \frac{V^\text{c}_0}{N}\frac{A}{(2\pi)^{2}} \int_{1/Na}^{1/a}\frac{dq}{q} \int_0^{2\pi} \frac{d\varphi}{\frac{\hbar^2 \cos^2\varphi}{2M_1}+\frac{\hbar^2 \sin^2\varphi}{2M_2}} = \frac{V^\text{c}_0}{N}\frac{A}{(2\pi)^{2}}  \frac{2\pi \ln L}{3 t a^2} J(t^\prime/t)
\end{align}
Here we introduced a dimensionless function 
\begin{align}
    J(t^\prime/t) = \int_0^{2\pi} \frac{d\varphi}{2\pi} \frac{6}{3(1+t^\prime / t)\cos^2\varphi+(1+5t^\prime / t) \sin^2\varphi} = \frac{2 \sqrt{3}}{\sqrt{5 (t^\prime / t) ^2+6 t^\prime / t +1}},\quad J(1)=1
\end{align}
The factor $A /N$ is the surface area of a unit cell of a triangular lattice, $s_0 = a^2 \sqrt{3}/2$. This finally gives,
\begin{align} 
    \frac{V^\text{c}_0}{t} \approx \frac{2\pi}{\ln L}\sqrt{5(t^{\prime}/t)^{2}+6t^{\prime}/t+1} \propto \frac{1}{\ln L}.
\end{align}
Importantly, this result shows that the critical value of the on-site term, $V^\text{c}_0$, tends to zero with the increasing lattice size. This signals that for an infinite system, the two bosons will form a bound state even at infinitesimal interaction strength.  

We illustrate the two boson localization in the Fig.\ref{figSup:TwoBoson} by solving the two-body problem numerically and looking into the average relative distance between the holes as a function of interaction strength $V_0$.  

\begin{figure}[h!]
\begin{centering}
\includegraphics[width=0.25\columnwidth]{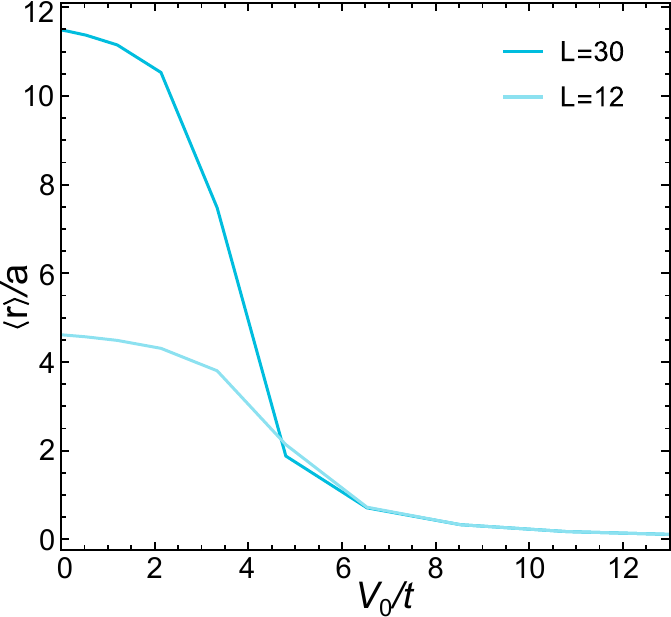}
\par\end{centering}
\caption{\textbf{Localization of two bosons}. Mean relative distance $\langle r\rangle /a$ between two bosons with $t^\prime / t = 1/2$ for $L=30, 12$ lattice sizes versus interaction strength $V_0/t$. For weak $V_0$ the relative distance is of order of the system size. With increasing $V_0$, the state tends to become localized.
}
\label{figSup:TwoBoson}
\end{figure}

\subsection{Three boson localization}

A key property that our models allows to study is the three-boson bound state. We perform an ED analysis with three bosons of different types. We start by looking into the average distance between the bosons in the GS. The calculations reveal that the relative distances between the bosons are equal and they decrease with increasing interaction $V_0$. The results are depicted in Fig. \ref{figSup:ThreeBoson}a) for two different system sizes, indicating the emergence of a three-boson bound state.

In order to determine which type of localization will take place in the case of finite doping, the two-boson or three-boson binding, we look into the binding energies per bosons in both cases. Let us first introduce the binding energies as 
\begin{align}
    \Delta^{(2)}(V_0) = E^{(2)}(V_0) - 2 E^{(1)},\quad \Delta^{(3)}(V_0) = E^{(3)}(V_0) - 3 E^{(1)},
\end{align}
for two and three particle states, respectively, and $E^{(1)} = -2(t+2t^\prime)$ is the single boson GS energy. And the binding energies per boson are then naturally defined as
\begin{align}
    \Delta^{(2)}_b = \Delta^{(2)}/2,\quad \Delta^{(3)}_b = \Delta^{(3)}/3.
\end{align}

We compare these values in the Fig. \ref{figSup:ThreeBoson}b). The results indicate that the binding per boson in the three-boson bound state is always stronger than in the two-boson bound state. This establishes, that the bosonic system in case of finite doping will condense into bosonic trios. Therefore, the 2H1M bosons will bound into 6H3M states with charge 6e and spin $S=6$.

\begin{figure}[h!]
\begin{centering}
\includegraphics[width=0.6\columnwidth]{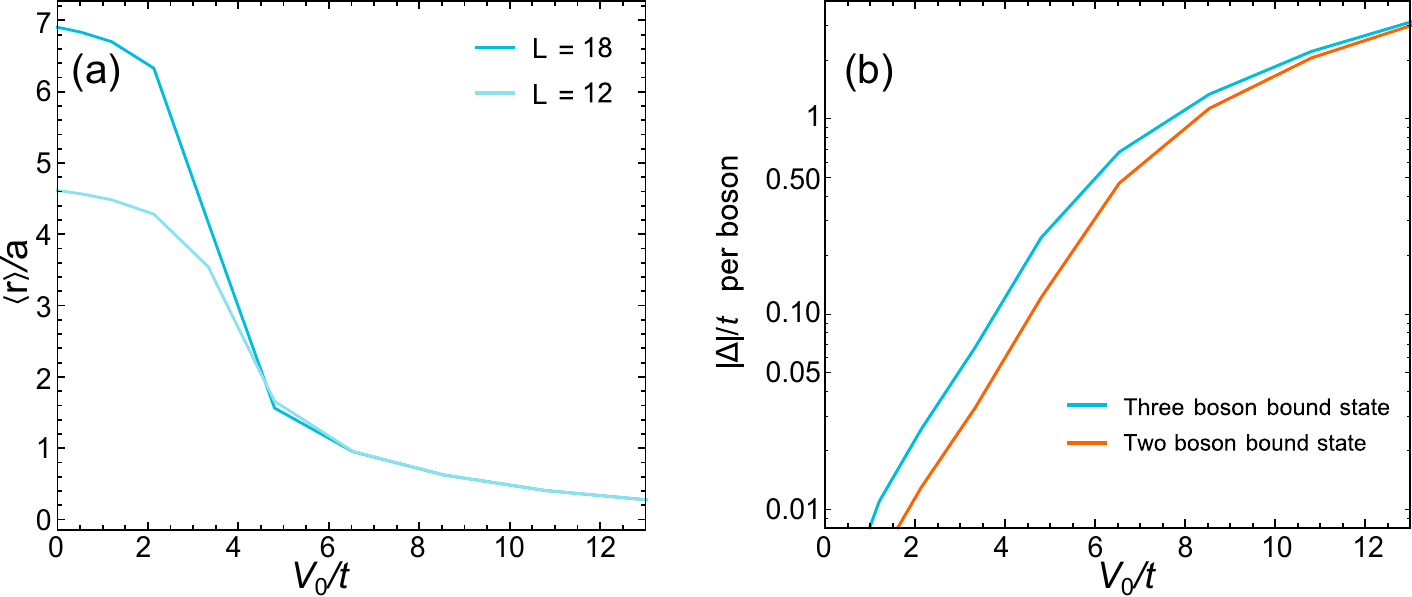}
\par\end{centering}
\caption{ \textbf{Localization of three bosons and comparison with two boson binding.} (a) Mean relative distance $\langle r\rangle /a$ between three bosons with $t^\prime / t = 1/2$ for $L=18, 12$ lattice linear sizes versus interaction strength $V_0/t$. For weak $V_0$ the relative distance is of order of the system size. With increasing $V_0$, the state tends to become localized. (b) The comparison of binding energies per boson for three-boson bound state $|\Delta^{(3)}_b|/t$ and two-boson bound state $|\Delta^{(2)}_b|/t$. The three-boson bound state is more favorable 
}
\label{figSup:ThreeBoson}
\end{figure}


\section{Hardcore boson model for 2H1M particles} 



In order to demonstate that the three-boson localization is a robust property, qualitatively independent on the model, here we present the analysis of another toy model leading to three 2H1M boson localization into a composite particles. We consider hard core bosons (forbidden double occupancy) with nearest neighbor attraction $V_0$. To obtain a lattice model with band minima at the $M$ points, we take into consideration a triangular lattice with both first and second nearest neighbor hoppings, similar to our main model. This model is written as
\begin{align}
    H= t_1^\prime \sum_{\langle \textbf{i}, \textbf{j} \rangle }\left(b_{ \textbf{i}}^{\dagger}b_{ \textbf{j}}+ h.c.\right)+t_2^\prime \sum_{\langle\langle \textbf{i}, \textbf{j} \rangle\rangle }\left(b_{ \textbf{i}}^{\dagger}b_{ \textbf{j}}+ h.c.\right) - V_0\sum_{ \langle \textbf{i}, \textbf{j} \rangle}n_{ \textbf{i}}n_{ \textbf{j}}, \quad t_2^\prime \geq 0.125 t_1^\prime
\end{align}

In this model, the effective masses of the particles near the band minima are anisotropic. We can determine them by expanding the dispersion around the $M_1$ point
\begin{align}
    m_x(M_1) = \frac{\hbar^2}{a^2 (9t^\prime_2 - t^\prime_1)}; \quad m_y(M_1) = \frac{\hbar^2}{3 a^2 (t^\prime_1 - t^\prime_2)};
\end{align}
This allows to tune the hopping parameters to obtain the actual effective masses of the 2H1S bosons.  

We then dope three bosons into the system and explore their interactions on a finite lattice of linear size $L$. In the absence of $V_0$, the bosons 
are delocalized. The Fig. \ref{figSup:hard_core_bosons} shows the localization of the bosons as a function of interaction strength $V_0$. We compare the average distance between the bosons for different system sizes of $L=12,18$. For $V_0\geq 0.3$, the both results agree with $20\%$ accuracy, providing a compelling evidence of three-particle localization. Finally, we look into the binding energy $\Delta$ of the bound state by comparing the energy of the composite boson to thrice that of a single boson at $M$ point. The inset to the Fig. \ref{figSup:hard_core_bosons} illustrates that while $\Delta$ is slightly positive for vanishing $V_0$, it already becomes negative for $V_0/t_1^\prime \sim 0.1$, indicating attraction.


\begin{figure}[h!]
\begin{centering}
\includegraphics[width=0.3\columnwidth]{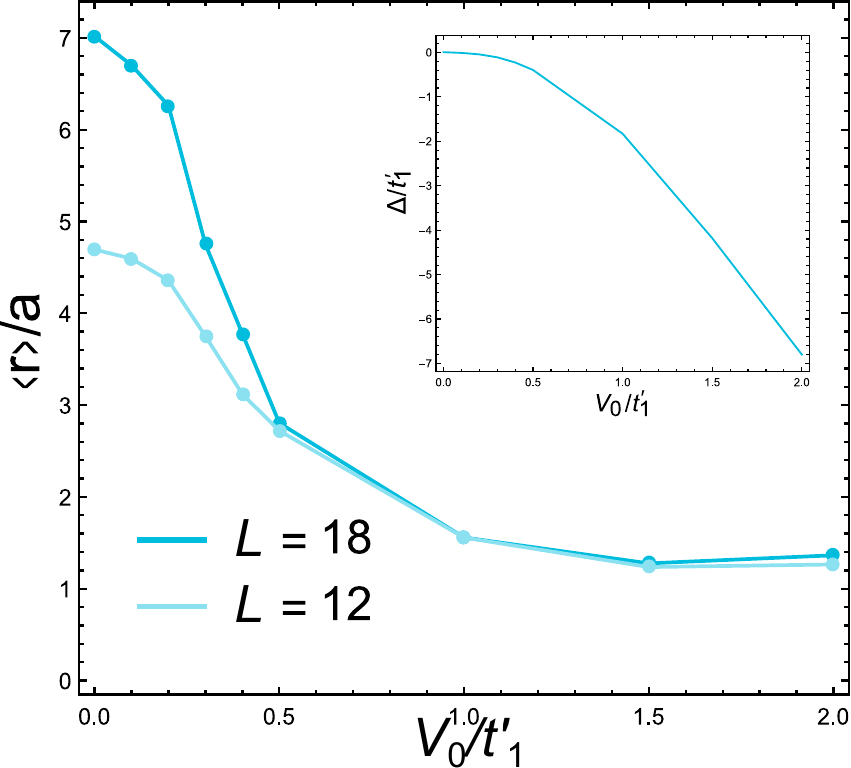}
\par\end{centering}
\caption{\textbf{Localization of hard-core bosons for $t^\prime_2=0.15 t^\prime_1$.} The figure illustrates the average distance between the bosons in the ground state as a function of $V_0$ for $L=12,18$. The inset shows the binding energy of the bound state.  
}
\label{figSup:hard_core_bosons}
\end{figure}

\newpage

\section{Magnonic Cooper Pairs in 1D}

In this section we extend the results of our initial model to show that the magnonic Cooper pairs emerge as a generic property of systems which have three-fold degenerate electron band maxima. We discuss a 1D chain with period boundary conditions with allowed nearest $(t_1)$, second-nearest $(t_2)$ and third-nearest neighbor $(t_3)$ hoppings. The electron dispersion in this model reads as
\begin{align}
E(k)=-2\left(t_{1}\cos\left(ka\right)+t_{2}\cos\left(2ka\right)+t_{3}\cos\left(3ka\right)\right)
\end{align}

For certain range of parameters $\left(t_{2},t_{3}\right)$ the band structure has three local maxima: one at $\pi/a$ and the two symmetric ones at some $\pm k_{0}$. The typical band structure in such system is depicted in Fig. \ref{figSup:1D_band_MCP}a). 
These maxima are exactly degenerate only at some line $t_{3}^{*}(t_{2})$ around which we expect to observe magnonic Cooper pair formation. 

\begin{figure}[h!]
\begin{centering}
\includegraphics[width=0.7\columnwidth]{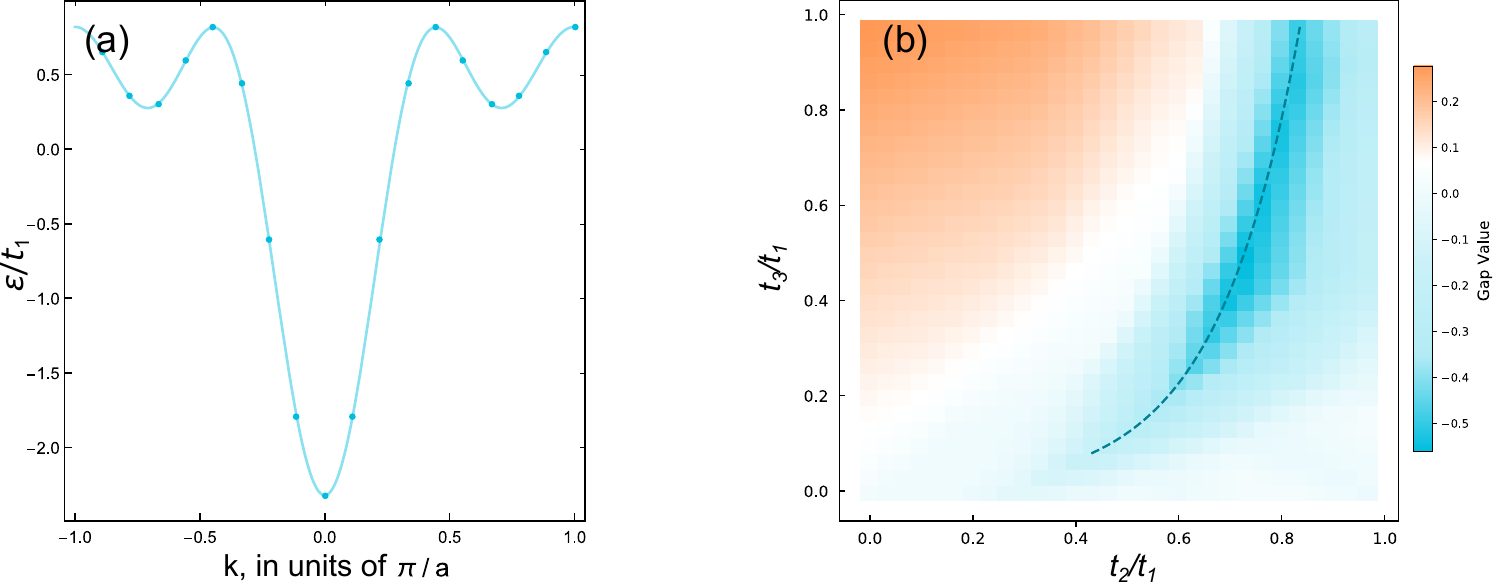}
\par\end{centering}
\caption{\textbf{Three-fold degenerate band maxima and formation of magnonic Cooper pairs in 1D.} (a) Band structure with three-fold degenerate maxima for $t_{2}/t_{1}=0.75, t_{3}/t_{1}=0.57$. Discrete points are calculated for N=18 sites. (b) The energy gap $\Delta/t_{1}$ in $t_{2}$-$t_{3}$ space calculated for $N=36$ sites. The region of the strongest attraction is at the dashed line $t_{3}^{*}(t_{2})$, which corresponds to the three-fold degeneracy of the electron band maxima (dashed line in Fig. \ref{figSup:1D_band_MCP}b)). 
}
\label{figSup:1D_band_MCP}
\end{figure}

To demonstrate the accuracy of this prediction, we compute the binding energy of 2H1M states compared to the 2H (two holes) and 1H+1H1M (hole and polaron) states in the infinite $U$ limit without magnetic field, similar to the approach used in the main text.
\begin{align}
    \Delta=E\left(2\text{H}1\text{M}\right)-\text{min}\left\{2E(1\text{H}), E(1\text{H}) + E(1\text{H}1M) \right\}
\end{align} 
In the Fig. \ref{figSup:1D_band_MCP}b), we see that the energy gap is positive far from the $t_{3}^{*}(t_{2})$ line, and then it becomes negative closer to the predicted region and peaks exactly at three-fold degeneracy line. Similarly to the triangular lattice calculations, there is a region where 1H+1H1M more favorable, but the 2H1M state becomes more favorable around the predicted line. This provides a strong evidence that the three-fold degeneracy is the key property leading to the magnonic Cooper pair formation.

We use this 1D system to study the interactions between several magnonic Cooper pairs. We choose $t_{2}/t_{1}=0.75,t_{3}/t_{1}=0.57$ and $N=18$ and compute the interaction energy between the 2H1M particle for increasing number of those particles, $n$.
\begin{align}
    \Delta\epsilon=E\left(n\times\text{2H1M}\right)-n\cdot E\left(\text{2H1M}\right)
\end{align} 
As discussed in the main text, we see that the interaction is repulsive and it increases with $n$. To explicitly illustrate the delocalization we look into the probability distribution as a function of relative distances between the magnons, for 2 and 3 magnonic Cooper pairs. As illustrated in the Fig. \ref{figSup:1D_band_MCP_repulsion}a),b), the wave function peaks when the distances between the particles is maximal in both cases with $L/2$ and $\pm L/3$ for 2 and 3 particles, respectively. Additionally, we look into the magnon-magnon correlation function for up to four 2H1M particles. The Fig. \ref{figSup:1D_band_MCP_repulsion}c) further evidences the repulsion between the particles.

\begin{figure}[h!]
\begin{centering}
\includegraphics[width=1\columnwidth]{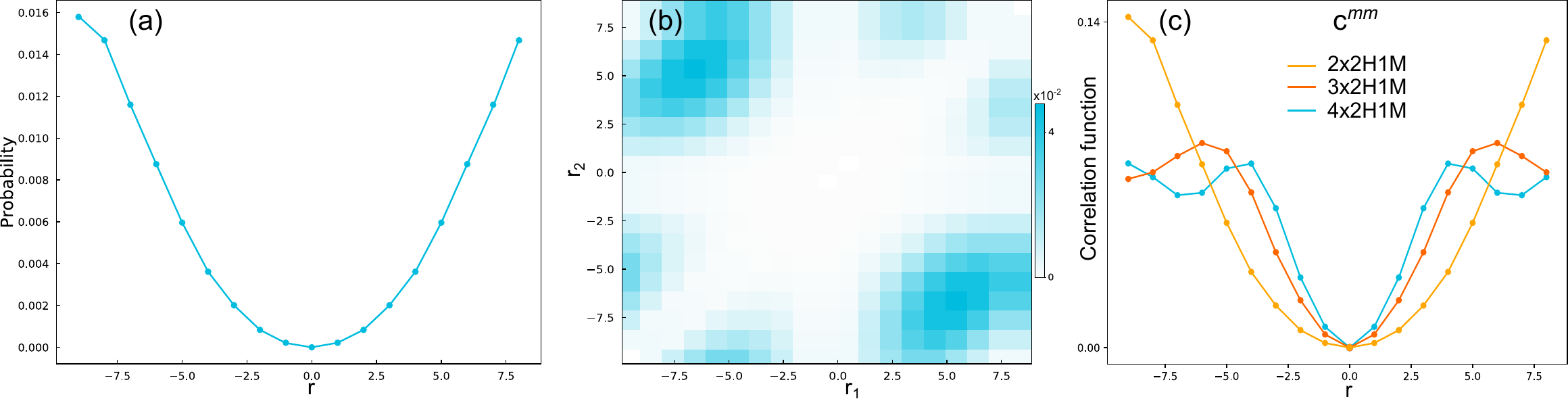}
\par\end{centering}
\caption{\textbf{Repulsion of several magnonic Cooper pair particles in 1D.} (a) Repulsion energy between the 2H1M particles. (b),(c) Probability distribution as a function of relative distances between magnons for two and three 2H1M particles, respectively. 
}
\label{figSup:1D_band_MCP_repulsion}
\end{figure}

\end{widetext}

\end{document}